\documentclass[journal,10pt]{IEEEtran}
\usepackage[cmex10]{amsmath}
\usepackage{graphicx}
\usepackage{color}
\usepackage{amsmath}
\usepackage{amssymb}
\usepackage{mathtools}
\usepackage{multicol}
\usepackage{subfigure}
\usepackage{blindtext}
\usepackage{algorithm}
\usepackage{algorithmic}
\usepackage{balance}
\usepackage{amsfonts}
\usepackage{bm}
\usepackage{stfloats}
\usepackage{subfig}
\usepackage{amsthm}
\usepackage{mathrsfs}
\usepackage{ulem}
\usepackage{CJK}

\usepackage{changepage}
\usepackage{cancel}
\usepackage{diagbox}
\usepackage{booktabs}
\usepackage{multirow}
\usepackage{float}
\usepackage{enumerate}
\usepackage{makecell}

\usepackage{cite}

\makeatletter

\newcommand{\Rmnum}[1]{\expandafter\@slowromancap\romannumeral #1@}
\makeatother

\ifCLASSINFOpdf
\else
\fi

\begin{document}

\title{Multi-objective Optimization of Space-Air-Ground Integrated Network Slicing Relying on a Pair of Central and Distributed Learning Algorithms}

\author{Guorong Zhou, Liqiang Zhao,~\IEEEmembership{Member,~IEEE},
Gan Zheng,~\IEEEmembership{Fellow,~IEEE},
Shenghui Song,~\IEEEmembership{Senior Member,~IEEE},
Jiankang Zhang,~\IEEEmembership{Senior Member,~IEEE},
and Lajos Hanzo,~\IEEEmembership{Life Fellow,~IEEE}

\thanks{This work was supported in part by the Key Research and Development Program of Shaanxi under Grant 2022KWZ-09, in part by the Postdoctoral Research Program of Shaanxi Province, in part by the Key-Area Research and Development Program of Guangdong Province under Grant 2020B0101120003, in part by the Fundamental Research Funds for the Central Universities under Grant QTZX23031, and in part by the 111 Project under Grant B08038.
The work of Shenghui Song was partially supported by a grant from the NSFC/RGC Joint Research Scheme sponsored by the Research Grants Council of the Hong Kong Special Administrative Region, China and National Natural Science Foundation of China (Project No. N\_HKUST656/22).
L. Hanzo would like to acknowledge the financial support of the Engineering and Physical Sciences Research Council projects EP/W016605/1, EP/X01228X/1 and EP/Y026721/1 as well as of the European Research Council's Advanced Fellow Grant QuantCom (Grant No. 789028).
\textit{(Corresponding author: Liqiang Zhao and Lajos Hanzo.)}
}

\thanks{Guorong Zhou is with the State Key Laboratory of Integrated Service Networks at Xidian University, Xi'an 710071, China (e-mail: guor\_zhou@163.com).}

\thanks{Liqiang Zhao is with the State Key Laboratory of Integrated Service Networks, Xidian University, Xi'an 710071, China, and also with the Guangzhou Institute of Technology, Xidian University, Guangzhou 510100, China (e-mail: lqzhao@mail.xidian.edu.cn).}

\thanks{Gan Zheng is with the School of Engineering, University of Warwick, Coventry, CV4 7AL, UK (e-mail: gan.zheng@warwick.ac.uk).}

\thanks{Shenghui Song is with the Department of Electronic and Computer Engineering, The Hong Kong University of Science and Technology, Hong Kong (e-mail: eeshsong@ust.hk).}

\thanks{Jiankang Zhang is with Department of Computing \& Informatics, Bournemouth University, BH12 5BB, UK (e-mail: jzhang3@bournemouth.ac.uk).}

\thanks{Lajos Hanzo is with the School of Electronics and Computer Science, University of Southampton, Southampton SO17 1BJ, UK (e-mail: lh@ecs.soton.ac.uk).}

\thanks{Copyright (c) 20xx IEEE. Personal use of this material is permitted. However, permission to use this material for any other purposes must be obtained from the IEEE by sending a request to pubs-permissions@ieee.org.}
}

{}

\maketitle

\begin{abstract}
As an attractive enabling technology for next-generation wireless communications, network slicing supports diverse customized services in the global space-air-ground integrated network (SAGIN) with diverse resource constraints.
In this paper, we dynamically consider three typical classes of radio access network (RAN) slices, namely high-throughput slices, low-delay slices and wide-coverage slices, under the same underlying physical SAGIN.
The throughput, the service delay and the coverage area of these three classes of RAN slices are jointly optimized in a non-scalar form by considering the distinct channel features and service advantages of the terrestrial, aerial and satellite components of SAGINs.
A joint central and distributed multi-agent deep deterministic policy gradient (CDMADDPG) algorithm is proposed for solving the above problem to obtain the Pareto optimal solutions.
The algorithm first determines the optimal virtual unmanned aerial vehicle (vUAV) positions and the inter-slice sub-channel and power sharing by relying on a centralized unit.
Then it optimizes the intra-slice sub-channel and power allocation, and the virtual base station (vBS)/vUAV/virtual low earth orbit (vLEO) satellite deployment in support of three classes of slices by three separate distributed units.
Simulation results verify that the proposed method approaches the Pareto-optimal exploitation of multiple RAN slices, and outperforms the benchmarkers.
\end{abstract}


\IEEEpeerreviewmaketitle

\section{Introduction}
Given the explosive proliferation of user equipment and service categories in the emerging sixth generation (6G) era, traditional terrestrial wireless communication networks are facing huge challenges \cite{1a,1}.
Thus, the space-air-ground integrated networking (SAGIN) \cite{2,3,4} concept was proposed for providing users with ubiquitous global coverage and enhanced data transmission.
Specifically, satellite networks are capable of providing seamless access to rural or remote areas, \textit{e.g.}, mountains, deserts and oceans.
As a complement, aerial networks \cite{50} are capable of providing prompt emergency communications in disaster areas, or providing service enhancement for areas having high tele-traffic density.
Finally, the densely deployed terrestrial networks can support high data rate transmission in urban areas.
By integrating the complementary advantages of the above three network components, SAGINs constitute a heterogeneous three-dimensional wireless communication architecture capable of providing seamless connections.

However, the variety of SAGIN-oriented service scenarios is increasing continually \cite{2,777}.
That means a diverse variety of different services share the underlying spectral resources, thus maintaining the target quality of service (QoS) of heterogeneous applications is hard to guarantee.
These services have distinct traffic features and may significantly affect each other, especially when the network resources are limited \cite{23}.
For example, bandwidth-extensive services are very likely to consume most of the channel resources, hence resulting in high transmission delays for lightweight delay-sensitive services.
Clearly, there is a paucity of research on how to support the service diversity encountered in SAGINs.
As an enabling technology of next-generation wireless networks, radio access network (RAN) slicing \cite{10,11} is potentially capable of constructing a set of independent virtual logical sub-networks based upon the same physical network infrastructure and resources, where each logical sub-network is customized for a particular type of service.
However, these network slices are isolated from each other.
Therefore, RAN slicing may be applied to support diverse customized services in SAGINs under various resource constraints \cite{2}.

To elaborate, RAN slicing is conceived for supporting various services, hence it is inappropriate to optimize a sliced network based on a single criterion \cite{11-1,11-2}.
For instance, bandwidth-extensive slicing aims for providing users with high-throughput services, while delay-sensitive slicing aims for ensuring ultra-low delay.
However, the objectives of maintaining a high throughput and ultra-low delay tend to conflict with each other.
To circumvent this conflict, multi-objective optimization problems (MOOPs) have been conceived as a mathematical framework for jointly optimizing multiple criteria \cite{12aa}, which are eminently suitable for RAN slicing scenarios.

There have been some studies on MOOPs in RAN slicing, but only in scalar form \cite{37,12,13,14,15}.
Although the traditional scalar method has low computational complexity, its objective function (OF) needs to be designed by experience, and its solution has a strong dependence on its weight.
A better way to solve MOOPs in RAN slicing is to use non-scalar methods, \textit{e.g.}, the popular multi-agent deep reinforcement learning (DRL) algorithm \cite{16,17}.
Instead of the unique globally optimal solution, the solutions of the non-scalar MOOP are composed of a number of non-dominated Pareto optimal solutions \cite{18}.
A multi-agent DRL algorithm is capable of learning different optimization objectives at the same time in the context of SAGINs, which makes it easier to obtain multiple Pareto optimal solutions \cite{20a, 20b}.

In this paper, we simultaneously establish three typical RAN slices, \textit{i.e.}, high-throughput slices, low-delay slices and wide-coverage slices, based upon the same underlying physical SAGIN.
A non-scalar MOOP is formulated for jointly optimizing the throughput, the service delay and the coverage area.
This is achieved by dynamically assigning the most appropriate network components, by finding the virtual unmanned aerial vehicles (vUAVs) having the optimal positions, and by allocating the most suitable subchannels and powers to the users relying on each slice type.
In order to solve the above non-scalar problem and to approach the Pareto optimal solutions, we propose a pair of central and distributed multi-agent deep deterministic policy gradient (CDMADDPG) algorithms.
More explicitly, our proposed method simultaneously optimizes three SAGIN slices, and finds near-Pareto optimal solutions.
Moreover, it succeeds in striking a compelling tradeoff among three types of RAN slices by exploiting the non-dominance relationship amongst the Pareto optimal solutions.
Against the above background, the main contributions of this paper are as follows:

\begin{itemize}
  \item We apply dynamic RAN slices in a global SAGIN scenario to support diverse customized services under specific resource constraints.
  The system model is built for jointly supporting the high-throughput slices, the low-delay slices and the wide-coverage slices based upon the same underlying 3-dimensional (3D) SAGIN.
  Then, the proposed model takes the distinct channel characteristics and service advantages of terrestrial, aerial and satellite communications in SAGIN into account, which is also verified by our simulation results.
  \item We formulate a non-scalar MOOP for jointly optimizing the throughput, the service delay and the coverage area based on RAN slicing, whilst the existing literature has only considered scalar MOOPs.
      To elaborate, we dynamically deploy virtual base stations (vBSs)/vUAVs/virtual low earth orbit (vLEO) satellites to support the users, and then find the optimal vUAVs' positions, allocate optimal subchannels and power resources among the slices, as well as optimize both the intra-slice subchannels and power resources for each user.
  \item We propose a CDMADDPG algorithm for solving the associated non-scalar MOOP.
  The algorithm first determines the position of the most suitable vUAVs and the associated inter-slice resource sharing, as decided by the centralized supervisory unit;
  then it proceeds to intra-slice resource sharing.
  This is arranged by the vBSs/vUAVs/vLEO satellites relying on three separate reduced-dimensional distributed units, in order to find near-Pareto optimal solutions.
  Our main contributions in comparison to the salient literature are summarized in Table~1.
\end{itemize}

\newcommand{\tabincell}[2]{\begin{tabular}{@{}#1@{}}#2\end{tabular}}
\begin{table*}[!ht]
\centering
\caption*{Table 1: Contrasting our contributions to the salient literature.}
\begin{tabular}{|l|c|c|c|c|c|c|c|c|}
  \hline
  Novelty & \cite{11-1}& \cite{11-2} & \cite{12} & \cite{12.1} & \cite{6} & \cite{19} & \cite{36a} & \tabincell{c}{Proposed} \\
  \hline
  Studying the SAGIN &  &  & & \checkmark & \checkmark &  & \checkmark & \checkmark \\
  \hline
  \tabincell{l}{Studying a dynamic sliced network} & \checkmark & \checkmark & \checkmark &  & & \checkmark &  & \checkmark \\
  \hline
  \tabincell{l}{Optimizing the vUAVs' positions} & & & & \checkmark & \checkmark &  &  & \checkmark \\
  \hline
  \tabincell{l}{Optimizing network components' assignment to users} &  &  & & \checkmark & \checkmark & \checkmark & \checkmark & \checkmark \\
  \hline
  \tabincell{l}{Optimization of inter-slice resources (\textit{e.g.}, power, subchannels) } & \checkmark & \checkmark &  &  & &  & & \checkmark \\
  \hline
  \tabincell{l}{Optimization of intra-slice resources between users (\textit{e.g.}, power, subchannels) } & \checkmark & \checkmark & \checkmark &  & \checkmark & & & \checkmark \\
  \hline
  \tabincell{l}{Jointly optimizing at least 3 metrics} &  &  &  &  &  & \checkmark & & \checkmark \\
  \hline
  \tabincell{l}{Formulating a non-scalar MOOP} &  &  &  &  &  & \checkmark & \checkmark & \checkmark \\
  \hline
  \tabincell{l}{\textbf{Solving the non-scalar MOOP by the multi-agent DRL algorithm}} & &  &  &  &  &  & & \textbf{\checkmark} \\
  \hline
  \tabincell{l}{\textbf{Solving the non-scalar MOOP by a pair of central and distributed algorithms}} &  &  &  & & & & & \textbf{\checkmark} \\
  \hline
\end{tabular}
\end{table*}

The rest of the paper is organized as follows.
In the next section, we briefly describe the related work of SAGIN and of MOOP based on RAN slicing.
In Section III, three different types of RAN slices of our SAGIN are established.
In Section IV, we formulate the proposed non-scalar MOOP of SAGIN slicing.
In Section V, we solve the MOOP by conceiving our CDMADDPG algorithm.
In Section VI, we discuss our simulation results.
Finally, the paper is concluded in Section VII.

\section{Related Work}
In this section, we first introduce the State of the Art in SAGIN, and then illustrate the popular methods of solving the MOOP based on RAN slicing.

\subsection{SAGIN}
SAGIN has been considered as one of the most promising techniques for increasing connectivity and improving the spatial 3D spectral efficiency of 6G system \cite{12.2}.
Substantial SAGIN research efforts have been focused on mobility management, traffic offloading and performance analysis \cite{5,12.1,6,7,8,9}.
For example,
Sun \textit{et al.} \cite{12.1} have presented an enhanced air-ground integrated vehicular network (AGVN) by introducing a surveillance plane for improving the management capability and have discussed the related networking, security, and application-oriented aspects.
Wang \textit{et al.} \cite{6} have proposed a data offloading scheme with drones acting as relays for maximizing the system capacity by jointly optimizing the users' connection scheduling, power control and drone trajectory.
An attractive transmission control strategy has been proposed in \cite{7} for optimally scheduling the ground-air-space and ground-to-space transmission schemes according to the specific status of ground users.
Mao \textit{et al.} \cite{8} have considered a space-aerial-assisted mixed cloud-edge computing offloading framework through jointly scheduling the associated position deployment and resource allocation.
Ye \textit{et al.} \cite{9} have constructed a cooperative channel model of SAGINs including the space-air, space-ground as well as air-ground links, and analyzed the outage performance attained.

However, while the variety of SAGIN-oriented service scenarios is increasing, the existing physical network architecture struggles to provide customized services for all users.
The next-generation enabling technology, RAN slicing, is capable of constructing a set of independent virtual logical sub-networks based upon the same physical network, with the promise of meeting the above challenge.
Nevertheless, at the time of writing, there are few studies on constructing network slices for SAGIN.

\subsection{MOOP based on RAN Slicing}
The key to solving a MOOP is to find Pareto optimal solutions, which are non-dominated by each other.
Generally, there are several methods available on RAN slicing, \textit{e.g.}, the traditional scalar method and the family of non-scalar intelligent algorithms.

The traditional scalar method mainly includes the weighted sum method, constraint method, etc \cite{37}.
This method aims to transform the MOOP into a single-objective optimization problem (SOOP), the optimal solution of which subjectively corresponds to one of the Pareto solutions.
Specifically, Shi \textit{et al.} \cite{12} have simultaneously analyzed the energy efficiency (EE) and delay in wireless network virtualization, but only the EE was employed as the objective, while the delay was treated as a constraint.
Afolabi \textit{et al.} \cite{13} have also used the constrained optimization method for minimizing the amount of computing resources subject to the constraint of a maximum tolerable mean response time on the end-to-end network slicing.
Although both the throughput and delay are considered in \cite{14}, the authors have used the weighted sum to project the above pair of metrics into a unified utility function.
Wang \textit{et al.} \cite{15} have integrated three optimization criteria into the slices' profits under a hybrid slicing framework.
However, these scalar methods have some drawbacks.
Explicitly, on the one hand, their OFs are subjectively designed based on experience, so it is hard to obtain a satisfactory solution in practice.
On the other hand, the optimization result of each slice is highly dependent on its weight in the OF, which requires extensive prior knowledge.

The non-scalar intelligent algorithm mainly includes the multi-objective evolutionary algorithms (MOEAs) \cite{16a} and the multi-agent DRL algorithms.
Specifically, MOEAs apply evolutionary operations to a population of legitimate candidates for finally approaching the Pareto front of optimal solutions.
For instance, Chantre \textit{et al.} \cite{19} employed MOEAs to solve the multi-criterion optimization problem built on 5G network slicing.
In the context of aeronautical ad hoc networks, Cui \textit{et al.} \cite{36a} formulated a twin-objective multi-hop routing problem and obtained multiple tradeoff solutions by MOEA.
Nevertheless, the existing literature only solved static and discrete MOOPs by employing MOEAs.

A multi-agent deep deterministic policy gradient (MADDPG) algorithm \cite{16} is capable of simultaneously learning different optimization objectives in a dynamic and successive SAGIN, but it may become computationally demanding in large-scale multi-user scenarios due to its single-layer and agent-coupled structure.
This will result in excessive complexity once the action set becomes excessively large in the training.
Hence the algorithm may fail to converge and the computational complexity becomes excessive.
Therefore we decompose the coupled structure of the MADDPG algorithm, and propose the CDMADDPG algorithm, where the heavy intra-slice resource decisions and training can be performed locally in the distributed units, and only their decision results have to be transmitted to the centralized supervisory unit for global processing.
Furthermore, by defining multiple agents within the distributed units as different types of RAN slices, the CDMADDPG algorithm succeeds in the simultaneous optimization of multiple objectives.

\section{System Model}
\begin{figure*}[!tp]
\setlength{\abovecaptionskip}{0cm}
\setlength{\belowcaptionskip}{0cm}
  \centering
  \includegraphics[width=7in]{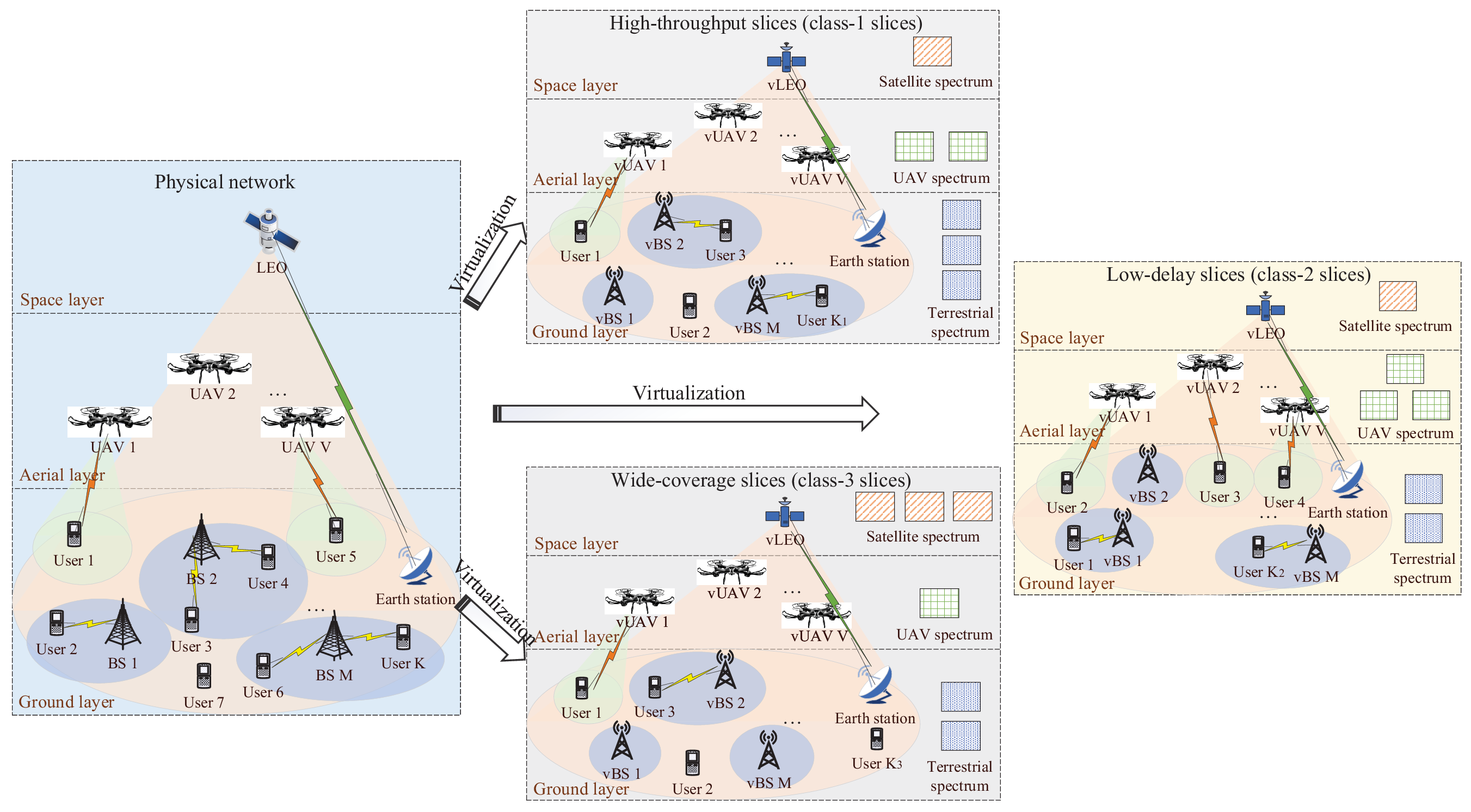}\\
  \caption{System model of SAGIN slicing.}\label{fig1}
\end{figure*}

Again, we focus our attention on analyzing three typical classes of RAN slices \cite{21} based upon the SAGIN of Fig.~\ref{fig1}.
Firstly, on the left side of Fig.~\ref{fig1}, we have a triple-layer physical RAN, constituted by the ground layer of $M$ base stations (BSs), the aerial layer of $V$ unmanned aerial vehicles (UAVs) and the space layer of a satellite constellation.
The UAVs are used as flying BSs \cite{26,27} and the satellite constellation is constituted by a group of low earth orbit (LEO) satellites to provide the full coverage of planet Earth.
The sets of $M$ and $V$ nodes are denoted by ${\bf{M}}\!=\! \{1, ... , M\}$ and ${\bf{V}}\!=\! \{1, ... , V\}$, respectively.
A total of $K$ terrestrial users may be served by different RAN slices, with the set of ${\bf{K}}\!=\! \{1, ... , K\}$.
We assume that all the terrestrial, airborne and spaceborne cellular downlink channels are operated in the C-band at 5 GHz, and the channels of different layers do not overlap with each other to avoid cross-layer interference \cite{32}.
The bandwidth of each layer is set to $B$ Hz, divided into $N$ subchannels associated with ${\bf{N}}\!=\! \{1, ... , N\}$, respectively.
Hence, each subchannel has a bandwidth of ${B/N}$ Hz, allowing the BSs/UAVs/LEOs to provide downlink communication for the users.
To reduce the model complexity, this paper does not consider the communication between network components (such as between the satellite and UAVs) \cite{23,8}.
The orthogonal frequency division multiple access (OFDMA) method is used for avoiding intra-cell interference.
We assume that the available power of each physical BS, UAV and LEO satellite is $P_B$, $P_V$ and $P_L$, respectively.

Then, with the aid of virtualization, we can support three typical classes of RAN slices, namely the high-throughput slices, the low-delay slices and the wide-coverage slices, which are referred to as class-$s$ slices, $s \!\in\! \left\{ {1,2,3} \right\}$, $S\!=\!3$.
As shown at the right of Fig.~\ref{fig1}, the network slicing process can be classified into three steps:
\begin{enumerate}
  \item All communication resources and components are dynamically shared among the three classes of slices. Every physical BS/UAV/LEO is mapped into $S$ vBSs/vUAVs/vLEOs, each of which is associated with a specific class-$s$ slice \cite{14}.
Naturally, the locations of the vBSs/vUAVs/vLEOs supporting the class-$s$ slices correspond to those of the physical BSs/UAVs/LEOs, thus we represent them by the same index $m,\!\forall m \!\in\! {\bf{M}}$ or $v,\!\forall v \!\in\! {\bf{V}}$ as those of the physical ones.
  \item Both the spectral and power resources of our 3D system are abstracted into the shared virtual resources.
  \item Depending on the different services' specific demands, the mobile network operator (MNO) orchestrates the flexible deployment of vBSs/vUAVs/vLEOs (also referred to as network components hereafter) as well as supports the real-time sharing and release of their associated communication resources among the different slices.
\end{enumerate}
We assume that ${K_s}$ users would like to request class-$s$ slices, with ${\bf{K_s}}\!=\! \{1, ... , K_s\}, {\bf{K_s}} \!\subset\! {\bf{K}}$.
A dynamically adaptive multi-slicing network is considered, where the time slots (TSs) are defined as the time interval $[t, t + 1)$, $t \in \{0, 1, 2, ..., T\!-\!1\}$, where $T$ is the total number of TSs and the TS duration is $\delta$ second.
In our model, it is assumed that the positions of vUAVs and users can change in a certain area from TS to TS, but they are stationary within a TS.
The users' positions follow a dynamic random distribution with a density ratio of 5:1 in densely populated and sparsely populated areas between TSs, respectively;
the vUAVs' positions will be dynamically optimized in the proposed optimization problem.\footnote{The users' and vUAVs' motion trajectory models are beyond the scope of this paper's discussions \cite{8}.} Naturally, the positions of vBSs remain fixed.
We also assume that the vLEO satellite index does not change within a total TS length $T$ and that a single satellite can be relied upon in our scenario \cite{6}.
Therefore, we denote the 3D coordinates of the vLEO satellite, vUAVs, terrestrial vBSs and users as $\left[ {x_{}^{vLEO}\!(t),y_{}^{vLEO}\!(t),z_{}^{vLEO}\!(t)} \right]$, $\left[ {x_v^{vUAV}\!(t),y_v^{vUAV}\!(t),z_v^{vUAV}\!(t)} \right]$, $\left[ x_m^{vBS},y_m^{vBS},z_m^{vBS} \right]$ and $\left[ {x_{k,s}^{UE}\!(t),y_{k,s}^{UE}\!(t),z_{k,s}^{UE}\!(t)} \right]$, respectively, where $z_m^{vBS} \!=\! z_{k,s}^{UE}\!(t) \!=\! 0$.
The main notations of this paper are summarized in Table 2.

\begin{table*}[!ht]
\centering
\caption*{Table 2: Main notations of this paper.}
\begin{tabular}{|l|r|}
  \hline
  \textbf{Notation} & \textbf{Description}  \\
   \hline
  ${\bf{M}}\!=\! \{1, ..., m, ..., M\}$ & \tabincell{r}{Set of vBSs}  \\
  \hline
  ${\bf{V}}\!=\! \{1, ... , v, ..., V\}$ & \tabincell{r}{Set of vUAVs}  \\
  \hline
  ${\bf{K}}\!=\! \{1, ... , k, ..., K\}$ & \tabincell{r}{Set of users}  \\
  \hline
  $P_B$, $P_V$, $P_L$ & \tabincell{r}{Available power of each physical BS/UAV/LEO satellite}  \\
  \hline
  $B$ & \tabincell{r}{System bandwidth}  \\
  \hline
  ${\bf{N}}\!=\! \{1, ... , n, ..., N\}$ & \tabincell{r}{Set of subchannels}  \\
  \hline
  ${\bf{K_s}}\!=\! \{1, ... , K_s\}, {\bf{K_s}} \!\subset\! {\bf{K}}$ & \tabincell{r}{Set of users requesting class-$s$ slices}  \\
  \hline
  $t$, $T$, $\delta$ & \tabincell{r}{TS; Total number of TSs; TS duration}  \\
  \hline
  ${g_{k,s,m,n}^{vBS}}(t)$, ${g_{k,s,v,n}^{vUAV}}(t)$, ${g_{k,s,n}^{vLEO}}(t)$ & \tabincell{r}{The channel coefficient of user $k$ on subchannel $n$ \\associated with vBS $m$/vUAV $n$/vLEO on class-$s$ slices at TS $t$}  \\
  \hline
  ${d_{k,s,m}^{vBS}}(t)$, ${\bf{d}}_{\bf{s}}^{{\bf{v}}{\bf{U}}{\bf{A}}{\bf{V}}} \!=\!\{{d_{k,s,v}^{vUAV}}(t)\}$, ${d_{k,s}^{vLEO}}(t)$ & \tabincell{r}{The distance between vBS $m$/vUAV $n$/vLEO and user $k$}  \\
  \hline
  ${{\bf{P}}_{\bf{s}}} \!=\! \left\{ {{\bf{P}}_{\bf{s}}^{{\bf{v}}{\bf{B}}{\bf{S}}}(t),{\bf{P}}_{\bf{s}}^{{\bf{v}}{\bf{U}}{\bf{A}}{\bf{V}}}(t),{\bf{P}}_{\bf{s}}^{{\bf{v}}{\bf{L}}{\bf{E}}{\bf{O}}}(t)} \right\}$ & \tabincell{r}{Set of transmit powers}  \\
  \hline
  ${{\bf{\Xi_s}}({t})\!=\!\{\xi_{k,s,n}(t)}\}$ & \tabincell{r}{Set of intra-slice subchannel allocation indicators}  \\
  \hline
  ${{\bf{\Phi_s}}({t})\!=\!\{{\varphi }_{k,s,m(v)}^{}\left( t \right)\}}$, ${\varphi }_{k,s,m(v)}^{}\left( t \right) \!\in\! \left\{ \varphi_{k,s,m}^{vBS} (t), \varphi_{k,s,v}^{vUAV} (t),\varphi_{k,s}^{vLEO} (t)\right\}$ & \tabincell{r}{Set of intra-slice network component association indicators}  \\
  \hline
  \tabincell{l}{$r_{k,s,m,n}^{vBS}\!\left( t \right)$, $R_{k,s}^{vBS}\!\left( t \right)$;  $r_{k,s,v,n}^{vUAV}\!\left( t \right)$, $R_{k,s}^{vUAV}\!\left( t \right)$;  $r_{k,s,n}^{vLEO}\!\left( t \right)$, $R_{k,s}^{vLEO}\!\left( t \right)$} & \tabincell{r}{Data rate and total data rate for user $k$ associated with \\ vBSs/vUAVs/vLEO satellite on class-$s$ slices}  \\
  \hline
  $I_{k,s,m,n}^{vBS}\!\left( t \right)$, $I_{k,s,v,n}^{vUAV}\!\left( t \right)$ & \tabincell{r}{Inter-cell interference of user $k$ associated with vBSs/vUAVs}  \\
  \hline
  $R_{k,s}\!\left( t \right)$, $R_1^{sum}\!\left( t \right)$ & \tabincell{r}{Total downlink data rate of user $k$ on class-$s$ slices;\\ Throughput of class-1 slices at TS $t$}  \\
  \hline
  $D_{k,2}\!\left( t \right)$, $D_2^{ave}\!\left( t \right)$ & \tabincell{r}{Service delay of user $k$, average delay of all users on class-2 slices}  \\
  \hline
  \tabincell{l}{${S\!I\!N\!R}_{k,3}^{vBS}\!\left( t \right)$, ${S\!I\!N\!R}_{k,3}^{vUAV}\!\left( t \right)$, ${S\!I\!N\!R}_{k,3}^{vLEO}\!\left( t \right)$, ${S\!I\!N\!R}_{3}^{ave}\!\left( t \right)$} & \tabincell{r}{SINR of user $k$ associated with vBSs/vUAVs/vLEO satellite \\on class-3 slices; average SINR on class-3 slices}  \\
  \hline
  ${P_{m,s}^{vBS}}$, ${P_{v,s}^{vUAV}}$, ${P_{s}^{vLEO}}$ & \tabincell{r}{Power consumption of each vBS/vUAV/vLEO satellite}  \\
  \hline
  $\beta$ & \tabincell{r}{Initial threshold of average delay objective}  \\
  \hline
  \tabincell{l}{${\bm{\rho _s}} \!=\! \left\{ {\rho _s^{vBS}\!\left( t \right),\!\rho _s^{vUAV}\! \left( t \right),\!\rho _s^{vLEO}\!\left( t \right)} \right\}$} & \tabincell{r}{Set of inter-slice power allocation factors}  \\
  \hline
   \tabincell{l}{${\bm{\eta _s}} \!=\! \left\{ {\eta _s^{vBS}\!\left( t \right),\!\eta _s^{vUAV}\!\left( t \right),\!\eta _s^{vLEO}\!\left( t \right)} \right\}$} & \tabincell{r}{Set of inter-slice subchannel allocation factors}  \\
  \hline
\end{tabular}
\end{table*}

\subsection{Terrestrial Communication}
In the terrestrial layer, it is assumed that the vBS provides customized services for multiple users.
The channel coefficient ${g_{k,s,m,n}^{vBS}}(t)$ of user ${k},\forall k \!\in\! {\bf{K_s}}$ on subchannel ${n}, \forall n \!\in\! {\bf{N}}$ associated with vBS ${m}$ on the class-$s$ slices at TS $t$ is independent and identically distributed (\textit{i.i.d.}) over time, which is modeled as:
\begin{equation} \label{1}
g_{k,s,m,n}^{vBS}(t) = h_{k,s,m,n}^{vBS}(t){\left[ {d_{k,s,m}^{vBS}(t)} \right]^{ - \alpha }},
\end{equation}
and
\begin{equation} \label{2}
d_{k,s,m}^{vBS}(t) \!=\! \left[ {{\left( {x_m^{vBS} \!-\! x_{k,s}^{UE}(t)} \right)}^2} \!+\! {{\left( {y_m^{vBS} \!-\! y_{k,s}^{UE}(t)} \right)}^2} \right]^{1/2},
\end{equation}
where $d_{k,s,m}^{vBS}(t)$ is the distance between the vBS $m$ and user $k$ of class-$s$ slices.
We adopt the classic Rayleigh fading channel model \cite{9,23}, and the channel fading of $h_{k,s,m,n}^{vBS}(t)$ follows an exponential distribution with unity mean,
while $\alpha$ is the path loss exponent.

Then, we define ${p_{k,s,m,n}^{vBS}}(t)$ as the power received from vBS ${m}$ by user ${k}$ associated with the subchannel ${n}$ on the class-$s$ slices at TS $t$.
Then we have ${\bf{P_s^{vBS}}}\!({t})\!=\!\{{p_{k,s,m,n}^{vBS}}\!(t)\}$.
Therefore, the data rate ${r_{k,s,m,n}^{vBS}(t)}$ of user ${k}$ on subchannel ${n}$ associated with vBS ${m}$ on class-$s$ slices at TS $t$ is given by
\begin{equation} \label{3}
r_{k,s,m,n}^{vBS}(t) = \frac{B}{N}{\log _2}\left[ {1 + \frac{{p_{k,s,m,n}^{vBS}(t)g_{k,s,m,n}^{vBS}(t)}}{{I_{k,s,m,n}^{vBS}\left( t \right) + (\frac{B}{N}){N_0}}}} \right],
\end{equation}
where
\begin{equation} \label{4}
I_{k,s,m,n}^{vBS}\left( t \right) = \sum _{j \in {\bf{M}}\backslash \left\{ {m} \right\}}^{}p_{k,s,j,n}^{vBS}(t)g_{k,s,j,n}^{vBS}(t).
\end{equation}
In \eqref{3}, ${{N_0}}$ is the power spectral density of the additive white Gaussian noise (AWGN) and $I_{k,s,m,n}^{vBS}\left( t \right)$ is defined as the inter-cell interference of user ${k}$ associated with vBS ${m}$ and subchannel ${n}$ on class-$s$ slices at TS $t$.
Moreover, ${g_{k,s,j,n}^{vBS}(t)}$ is the channel coefficient of the interfering vBS ${j},\forall j \!\in \!{\bf{M}}\!\backslash\!\{m\}$;
and ${p_{k,s,j,n}^{vBS}(t)}$ is the interference power of vBS ${j}$.

\subsection{UAV Communication}
In the aerial layer, UAVs can furnish users with various types of slices through their downlink channels.
Similarly, the channel coefficient ${g_{k,s,v,n}^{vUAV}}(t)$ of user ${k}$ on subchannel ${n}$ associated with vUAV ${v}$ on class-$s$ slices at TS $t$ is also \textit{i.i.d.} vs. time, formulated as follows:
\begin{equation} \label{8}
\begin{split}
&g_{k,s,v,n}^{vUAV}(t) = h_{k,s,v,n}^{vUAV}(t){\left[ {d_{k,s,v}^{vUAV}(t)} \right]^{ - \alpha }}\\
& \!=\! {h_0}{\left[ {d_{k,s,v}^{vUAV}(t)} \right]^{ - \alpha }}\left( {\frac{R}{{R \!+\! 1}} \hat{h}_{k,s,v,n}^{vUAV}(t) \!+\! \frac{1}{{R \!+\! 1}} \tilde{h}_{k,s,v,n}^{vUAV}(t)} \right)\!,\!
\end{split}
\end{equation}
and
\begin{equation} \label{9}
\begin{split}
&d_{k,s,v}^{vUAV}(t) \!=\! \left[ {\left( {x_v^{vUAV}(t) \!-\! x_{k,s}^{UE}(t)} \right)^2} \right.\\
&\left. + {\left( {y_v^{vUAV}(t) \!-\! y_{k,s}^{UE}(t)} \right)^2} \!+\! {\left( {z_v^{vUAV}(t)} \right)^2} \right]^{1/2},
\end{split}
\end{equation}
where $d_{k,s,v}^{vUAV}(t)$ is the distance between the vUAV $v$ and user $k$ relying on class-$s$ slices.
We have ${\bf{d}}_{\bf{s}}^{{\bf{v}}{\bf{U}}{\bf{A}}{\bf{V}}} \!=\! \left\{ {d_{k,s,v}^{vUAV}(t)} \right\}$ as one of the decision variables, because it dynamically changes between TSs.
The fading of $h_{k,s,v,n}^{vUAV}(t)$ is assumed to follow a Rician channel model \cite{8,31}, in which $h_0$ denotes the reference channel gain when the distance is 1 meter;
$R$ is the Rician fading factor;
$\hat{h}_{k,s,v,n}^{vUAV}(t)$ represents the line-of-sight (LoS) component that satisfies $\left| {\hat{h}_{k,s,v,n}^{vUAV}(t)} \right| \!=\! 1$;
and $\tilde{h}_{k,s,v,n}^{vUAV}(t)$ is the non-line-of-sight (NLoS) component that follows $\tilde{h}_{k,s,v,n}^{vUAV}(t)\sim \mathcal{CN}\left( {0,1} \right)$.

Furthermore, ${p_{k,s,v,n}^{vUAV}}(t)$ is the power transmitted from vUAV ${v}$ to user ${k}$ associated with subchannel ${n}$
on class-$s$ slices at TS $t$, with the set as ${\bf{P_s^{vUAV}}}({t})\!=\!\{{p_{k,s,v,n}^{ vUAV }}(t)\}$.
The data rate ${r_{k,s,v,n}^{ vUAV }(t)}$ of user ${k}$ on subchannel ${n}$ associated with vUAV ${v}$ on class-$s$ slices at TS $t$ is given by
\begin{equation} \label{10}
r_{k,s,v,n}^{vUAV}(t) = \frac{B}{N}{\log _2}\left[ {1 + \frac{{p_{k,s,v,n}^{vUAV}(t)g_{k,s,v,n}^{vUAV}(t)}}{{I_{k,s,v,n}^{vUAV}\left( t \right) + (\frac{B}{N}){N_0}}}} \right],
\end{equation}
where
\begin{equation} \label{11}
I_{k,s,v,n}^{vUAV}\left( t \right) = \sum _{i \in {\bf{V}}\backslash \left\{ {v} \right\}}^{}p_{k,s,i,n}^{vUAV}(t)g_{k,s,i,n}^{vUAV}(t).
\end{equation}
Similarly, $I_{k,s,v,n}^{vUAV}\left( t \right)$ is the inter-cell interference of user ${k}$ associated with vUAV ${v}$ and subchannel ${n}$ on class-$s$ slices at TS $t$,
where ${g_{k,s,i,n}^{vUAV}(t)}$ and ${p_{k,s,i,n}^{vUAV}(t)}$
are the channel coefficient and the power of the interference vUAV ${i},\forall i \!\in \!{\bf{V}}\!\backslash\!\{v\}$, respectively.

\subsection{Satellite Communication}
With the aid of LEO satellite downlink channels, seamless coverage may be provided for users in our slicing network.
In the space layer, the channel coefficient ${g_{k,s,n}^{vLEO}}(t)$ of user ${k}$ on subchannel ${n}$ associated with the vLEO satellite on class-$s$ slices at TS $t$ is defined as:
\begin{equation} \label{14}
\begin{split}
&g_{k,s,n}^{vLEO}(t) = h_{k,s,n}^{vLEO}(t){\left[ {d_{k,s}^{vLEO}(t)} \right]^{ - \alpha }}\\
& = {\left( {\frac{c}{{4\pi {f_c}}}} \right)^2}{\left[ {d_{k,s}^{vLEO}(t)} \right]^{ - \alpha }},
\end{split}
\end{equation}
and
\begin{equation} \label{14b}
\begin{split}
&d_{k,s}^{vLEO}(t) \!=\! \left[ {\left( {x^{vLEO}(t) \!-\! x_{k,s}^{UE}(t)} \right)^2} \right. \\
& \left. + {\left( {y^{vLEO}(t) \!-\! y_{k,s}^{UE}(t)} \right)^2} \!+\! {\left( {z^{vLEO}(t)} \right)^2} \right]^{1/2},
\end{split}
\end{equation}
where $d_{k,s}^{vLEO}(t)$ is the distance between the vLEO satellite and user $k$ of class-$s$ slices.
Furthermore, $h_{k,s,n}^{vLEO}(t)$ is the unit radio propagation loss of the satellite link caused by the free space loss \cite{32},
where ${f_c}$ is the carrier frequency and $c$ is the velocity of light.

We denote the transmit power of the vLEO satellite to user ${k}$ associated with subchannel ${n}$ on class-$s$ slices at TS $t$ by ${p_{k,s,n}^{vLEO}}(t)$, having ${\bf{P_s^{vLEO}}}({t})\!=\!\{{p_{k,s,n}^{ vLEO }}(t)\}$.
Thus, the data rate ${r_{k,s,n}^{ vLEO }(t)}$ of user ${k}$ on subchannel ${n}$ associated with the vLEO satellite on class-$s$ slices at TS $t$ is given by
\begin{equation} \label{15}
r_{k,s,n}^{vLEO}(t) = \frac{B}{N}{\log _2}\left[ {1 + \frac{{p_{k,s,n}^{vLEO}(t)g_{k,s,n}^{vLEO}(t)}}{{(\frac{B}{N}){N_0}}}} \right].
\end{equation}

Subsequently, we define ${\xi_{k,s,n}(t)}$ as a binary factor of user $k$ associated with subchannel ${n}$ on class-$s$ slices at TS $t$, and ${{\bf{\Xi_s}}({t})\!=\!\{\xi_{k,s,n}(t)}\}$ is the set of subchannel allocation indicators.
Furthermore, ${{\bf{\Phi_s}}({t})\!=\!\{{\varphi }_{k,s,m(v)}^{}\left( t \right)\}}$ represents the set of network component association indicators, where ${\varphi }_{k,s,m(v)}^{}\left( t \right) \!\in\! \left\{ \varphi_{k,s,m}^{vBS} (t), \varphi_{k,s,v}^{vUAV} (t),\varphi_{k,s}^{vLEO} (t)\right\}$ is a binary factor of user $k$ associated with vBS ${m}$, vUAV $v$ or the vLEO satellite on class-$s$ slices at TS $t$.
Their specific expressions are as follows:
\begin{equation} \label{5}
{\xi_{k,s,n}\!\left( t \right) \!=\! \left\{ \begin{split}
&1,\mbox{if\;}n\mbox{\;is\;allocated\;to\;user\;}k\mbox{\;of\;class-}s\mbox{\;slices},\!\\
&0,\mbox{otherwise},
\end{split} \right.}
\end{equation}
and
\begin{equation} \label{6}
{ {\varphi }_{k,s,m(v)}^{}\!\left( t \right) \!=\! \left\{ \begin{split}
&1, \mbox{if\;user\;}k\mbox{\;of\;class-}s\mbox{\;slices\;is\;associated}\!\\
&~~\mbox{\;with\;}m,v\mbox{\;or\;vLEO},\!\\
&0, \mbox{otherwise}.
\end{split} \right.}
\end{equation}

Hence, the downlink data rate of user $k$ associated with vBSs, vUAVs and the vLEO satellite on class-$s$ slices at TS $t$ are formulated as, respectively:
\begin{equation} \label{7}
R_{k,s}^{vBS}\left( t \right) = \sum\limits_{n \in {\bf{N}}}^{} {\sum\limits_{m \in {\bf{M}}}^{} {\varphi _{k,s,m}^{vBS}(t)\xi _{k,s,n}^{}(t)r_{k,s,m,n}^{vBS}(t)} },
\end{equation}
\begin{equation} \label{13}
R_{k,s}^{vUAV}\left( t \right) = \sum\limits_{n \in {\bf{N}}}^{} {\sum\limits_{v \in {\bf{V}}}^{} {\varphi _{k,s,v}^{vUAV}(t)\xi _{k,s,n}^{}(t)r_{k,s,v,n}^{vUAV}(t)} },
\end{equation}
and
\begin{equation} \label{17}
R_{k,s}^{vLEO}\left( t \right) = \sum\limits_{n \in {\bf{N}}}^{} {\varphi _{k,s}^{vLEO}(t)\xi _{k,s,n}^{}(t)r_{k,s,n}^{vLEO}(t)}.
\end{equation}

Upon considering our triple-layer RAN slicing system including the space, air and ground layers, we can get the total downlink data rate of user $k$ on class-$s$ slices at TS $t$, as follows:
\begin{equation} \label{18}
R_{k,s}^{}\left( t \right) = R_{k,s}^{vBS}\left( t \right) + R_{k,s}^{vUAV}\left( t \right) + R_{k,s}^{vLEO}\left( t \right).
\end{equation}

\subsection{SAGIN Slices}
Let us now build the customized optimization objectives of three classes of slices, respectively.
As for class-1 slices, such as a high-resolution video streaming service, MNO has to satisfy the users' requirements by transmitting high-throughput data through the wireless downlink.
Hence, the throughput of the system is regarded as the customized optimization objective for class-1 slices.
We express the throughput of class-1 slices at TS $t$ as:
\begin{equation} \label{19}
R_1^{sum}\left( t \right) = \sum\limits_{k \in {{\bf{K}}_{\bf{1}}}}^{} {{R_{k,1}}\left( t \right)} .
\end{equation}

As for class-2 slices, such as mission critical services and emergency response, the primary target is to reduce the service delay in order to guarantee flawless lip-synchronization.
We represent by $A_{2}\!\left( t \right)\!=\!\{A_{1,2}\!\left( t \right)\!,\!A_{2,2}\!\left( t \right)\!,\!...,\!A_{k,2}\!\left( t \right)\!,\!...\}$ the process of random data arrivals on class-2 slices at TS $t$, which follows a Poisson arrival process having the average arrival rate of $\lambda_2$.
In this context, $A_{2}\left( t \right)$ is assumed to be independent among the users and \textit{i.i.d.} across the TSs.
Consequently, the service procedure at TS $t$ can be modeled as an M/D/1 queue \cite{24,25} having the service rate of ${R_{k,2}}\left( t \right)$.
The service delay of user $k$ on class-2 slices at TS $t$ is denoted by ${D_{k,2}}\left( t \right)$, which consists of the propagation delay, the transmission delay and the queuing delay, yielding:
\begin{equation} \label{20}
{D_{k,2}}\left( t \right) \!=\! \frac{{{d_{k,2}}\left( t \right)}}{c} \!+\! \frac{{{A_{k,2}}\left( t \right)}}{{{R_{k,2}}\left( t \right)}} \!+\! \frac{{{\lambda _2}{A_{k,2}}\left( t \right)}}{{2\left[ {{{\left( {R_{k,2}^{}\left( t \right)} \right)}^2} \!-\! {\lambda _2}{R_{k,2}}\left( t \right)} \right]}}\!,\!
\end{equation}
where we have:
\begin{equation} \label{21}
\begin{split}
&{d_{k,2}}\left( t \right) \!=\! \sum\limits_{n \in {\bf{N}}}^{} {\sum\limits_{m \in {\bf{M}}}^{} {\varphi _{k,2,m}^{vBS}(t)\xi _{k,2,n}^{}(t)d_{k,2,m}^{vBS}(t)} } \\
& \!+\! \sum\limits_{n \in {\bf{N}}}^{} {\sum\limits_{v \in {\bf{V}}}^{} {\varphi _{k,2,v}^{vUAV}(t)\xi _{k,2,n}^{}(t)d_{k,2,v}^{vUAV}(t)} }  \\
&\!+\! \sum\limits_{n \in {\bf{N}}}^{} {\varphi _{k,2}^{vLEO}(t)\xi _{k,2,n}^{}(t)d_{k,2}^{vLEO}(t)} .
\end{split}
\end{equation}
The first term of \eqref{20} is the propagation delay of user $k$ on class-2 slices at TS $t$, where ${d_{k,2}}\left( t \right)$ is the distance between the user $k$ on class-2 slices and its associated network component.
The second term of \eqref{20} is the transmission delay of user $k$ on class-2 slices at TS $t$, and the last term of \eqref{20} is the queuing delay of user $k$ on class-2 slices at TS $t$.

The average delay of all users on class-2 slices can be expressed as
\begin{equation} \label{22}
\setlength{\abovedisplayskip}{2pt}
\setlength{\belowdisplayskip}{2pt}
D_2^{ave}\left( t \right) = \frac{1}{{{K_2}}}\sum\limits_{k \in {\bf{K_2}}}^{} {{D_{k,2}}\left( t \right)}.
\end{equation}
In order to meet the low-delay requirements of users on class-2 slices, we aim for minimizing $D_2^{ave}\left( t \right)$.

Class-3 slices provide the basic wide-area access services at a low user density, in rural areas supporting few people.
Given the time-varying channel state and the heterogeneous 3D network architecture, the basic access services are also of variable-rate nature.
Consequently, the main optimization objective of class-3 slices is to maximize the average signal to interference plus noise power ratio (SINR) of the network for ensuring a wide coverage area.
The SINR of user $k$ associated with vBSs, vUAVs and vLEO satellite on class-3 slices at TS $t$ are defined respectively as:
\begin{equation} \label{23}
\setlength{\abovedisplayskip}{2pt}
\setlength{\belowdisplayskip}{2pt}
\begin{split}
&\!S\!I\!N\!R_{k,3}^{vBS}(t) \\
&\!=\! \sum\limits_{n \in {\bf{N}}}^{} {\sum\limits_{m \in {\bf{M}}}^{} {\varphi _{k,3,m}^{vBS}(t)\xi _{k,3,n}^{}(t)\frac{{p_{k,3,m,n}^{vBS}(t)g_{k,3,m,n}^{vBS}(t)}}{{I_{k,3,m,n}^{vBS}(t) + (\frac{B}{N}){N_0}}}} } ,
\end{split}
\end{equation}
\begin{equation} \label{24}
\setlength{\abovedisplayskip}{2pt}
\setlength{\belowdisplayskip}{2pt}
\begin{split}
&\!S\!I\!N\!R_{k,3}^{vUAV}(t) \\
&\!=\!\sum\limits_{n \in {\bf{N}}}^{} {\sum\limits_{v \in {\bf{V}}}^{} {\varphi _{k,3,v}^{vUAV}(t)\xi _{k,3,n}^{}(t)\frac{{p_{k,3,v,n}^{vUAV}(t)g_{k,3,v,n}^{vUAV}(t)}}{{I_{k,3,v,n}^{vUAV}(t) + (\frac{B}{N}){N_0}}}} } ,
\end{split}
\end{equation}
and
\begin{equation} \label{25}
\setlength{\abovedisplayskip}{2pt}
\setlength{\belowdisplayskip}{2pt}
\!S\!I\!N\!R_{k,3}^{vLEO}(t) \!=\! \sum\limits_{n \in {\bf{N}}}^{} {\varphi _{k,3}^{vLEO}(t)\xi _{k,3,n}^{}(t)\frac{{p_{k,3,n}^{vLEO}(t)g_{k,3,n}^{vLEO}(t)}}{{(\frac{B}{N}){N_0}}}} .
\end{equation}

By jointly considering our SAGIN slicing network,
we express the average SINR of class-3 slices as:
\begin{equation} \label{27}
\setlength{\abovedisplayskip}{2pt}
\setlength{\belowdisplayskip}{2pt}
\begin{split}
&\!S\!I\!N\!R_3^{ave}\left( t \right) \!=\! \frac{1}{{{K_3}}}\sum\limits_{k \in {\bf{K_3}}}^{} \left\{\!S\!I\!N\!R_{k,3}^{v\!B\!S}\left( t \right) \!+ \!S\!I\!N\!R_{k,3}^{v\!U\!A\!V}\left( t \right) \right.\\
&\left.\!+ \!S\!I\!N\!R_{k,3}^{v\!L\!E\!O}\left( t \right)\right\}.
\end{split}
\end{equation}

\vspace{-0.05in}
\section{Multi-Objective Problem Formulation}
\begin{figure}[!tp]
\setlength{\abovecaptionskip}{0cm}
\setlength{\belowcaptionskip}{0cm}
  \centering
  \includegraphics[width=3in]{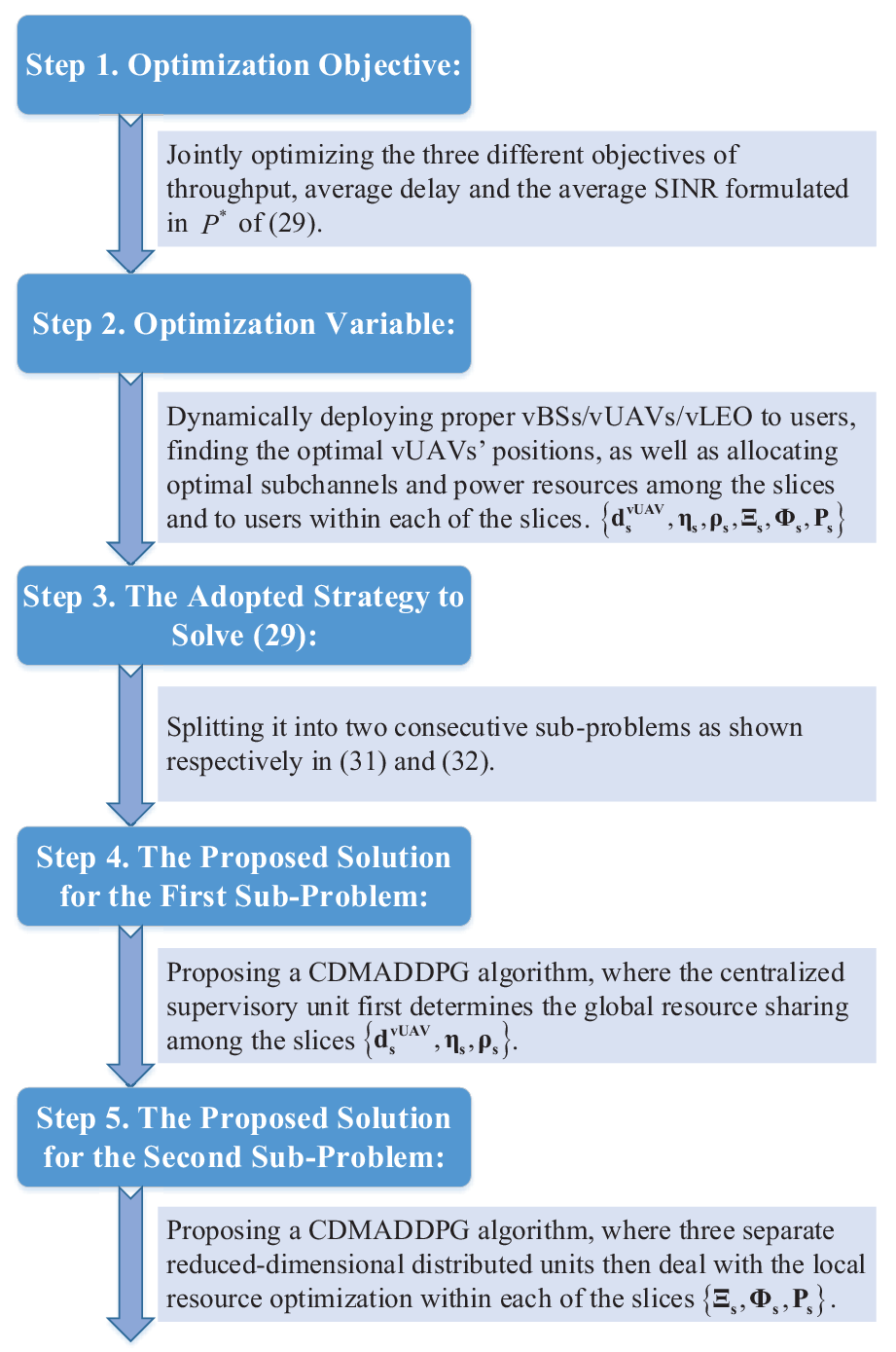}\\
  \caption{Flow of the mathematical analysis.}\label{fig0}
\end{figure}

Before starting this section, we provide Fig.~\ref{fig0} to emphasize the flow of the mathematical analysis in the sequel. This diagram helps readers to understand the flow of the paper.

Explicitly, we formulate the power consumption of each vBS, vUAV and vLEO satellite on class-$s$ slices respectively as:
\begin{equation} \label{28}
\setlength{\abovedisplayskip}{2pt}
\setlength{\belowdisplayskip}{2pt}
P_{m,s}^{vBS}\left( t \right) \!=\! \sum\limits_{n \in {\bf{N}}}^{} \! \sum\limits_{k \in {{\bf{K}}_{\bf{s}}}}^{} \! \varphi _{k,s,m}^{vBS}(t)\xi _{k,s,n}^{}(t)p_{k,s,m,n}^{vBS}(t),
\end{equation}
\begin{equation} \label{29}
\setlength{\abovedisplayskip}{2pt}
\setlength{\belowdisplayskip}{2pt}
P_{v,s}^{vUAV}\left( t \right) \!=\! \sum\limits_{n \in {\bf{N}}}^{} \! \sum\limits_{k \in {{\bf{K}}_{\bf{s}}}}^{}  \!\varphi _{k,s,v}^{vUAV}(t)\xi _{k,s,n}^{}(t)p_{k,s,v,n}^{vUAV}(t),
\end{equation}
and
\begin{equation} \label{30}
\setlength{\abovedisplayskip}{2pt}
\setlength{\belowdisplayskip}{2pt}
P_{s}^{vLEO}\left( t \right) \!=\! \sum\limits_{n \in {\bf{N}}}^{} \! \sum\limits_{k \in {{\bf{K}}_{\bf{s}}}}^{} \! \varphi _{k,s}^{vLEO}(t)\xi _{k,s,n}^{}(t)p_{k,s,n}^{vLEO}(t).
\end{equation}

Next, we define the set of inter-slice power allocation factors of the vBSs, vUAVs and vLEO satellite in class-$s$ slices at TS $t$ as ${\bm{\rho _s}} \!=\! \left\{ {\rho _s^{vBS}\!\left( t \right),\!\rho _s^{vUAV}\! \left( t \right),\!\rho _s^{vLEO}\!\left( t \right)} \right\}$.
Furthermore, ${\bm{\eta _s}} \!=\! \left\{ {\eta _s^{vBS}\!\left( t \right),\!\eta _s^{vUAV}\!\left( t \right),\!\eta _s^{vLEO}\!\left( t \right)} \right\}$ represents the set of inter-slice subchannel allocation factors of the vBSs, vUAVs and vLEO satellite in class-$s$ slices at TS $t$.

As for our dynamically adaptive multi-slicing SAGIN, it is more effective to adopt the long-term objectives to estimate the network performance instead of optimizing near-instantaneous objectives \cite{32c}.
By \eqref{19}, \eqref{22} and \eqref{27}, we can jointly optimize the three different objectives of throughput, average delay and the average SINR relying on multi-objective optimization (MOO) algorithms in the non-scalar form, which is shown in ${\bf{P}}^*$ of \eqref{31} at the top of the next page.
\begin{figure*}[ht]
\begin{equation}\label{31}
\setlength{\abovedisplayskip}{2pt}
\setlength{\belowdisplayskip}{2pt}
\begin{aligned}
&{{\bf{P}}^*}\!:\! \mathop {\max }\limits_{\left\{ {\scriptstyle{{\bf{\Xi }}_{\bf{s}}}{\bf{,}}{{\bf{\Phi }}_{\bf{s}}}{\bf{,}}{{\bf{P}}_{\bf{s}}},\hfill\atop
\scriptstyle{\bm{\eta _s}}\!,{\bm{\rho_s}}{\bf{,d}}_{\bf{s}}^{{\bf{\!v\!U\!A\!V\!}}}\hfill} \right\}} \!
\left\{ \!\begin{split}
&\sum\nolimits_{t =\! 1}^T {} R_1^{sum}\!\left( t \right)\\
&\sum\nolimits_{t =\! 1}^T {} \left[ {\beta  \!-\! D_2^{ave}\!\left( t \right)} \right]\\
&\sum\nolimits_{t =\! 1}^T {} \!S\!I\!N\!R_3^{ave}\!\left( t \right)
\end{split} \!\right\},\\
&\text{s.t.}
~ C1: \sum\limits_{k \in {{\bf{K}}_s}}^{} {\xi _{k,s,n}^{}\left( t \right){\varphi}_{k,s,m(v)}^{}\!\left( t \right) \!\le\! 1} ,\forall n \!\in\! {\bf{N}},m \!\in\! {\bf{M}},v \!\in\! {\bf{V}}\!,\!\\
&~~~~ C2:\sum\limits_{n \in {\bf{N}}}^{} {\sum\limits_{\scriptstyle m\in {\bf{M}}\hfill\atop
\scriptstyle (v\in {\bf{V}})\hfill}^{} {\xi _{k,s,n}^{}\left( t \right){\varphi}_{k,s,m(v)}^{}\left( t \right) \le 1} } ,\forall k \in {{\bf{K}}},\\
&~~~~ C3:\sum\limits_{m \in {\bf{M}}}^{} {\varphi _{k,s,m}^{vBS}\left( t \right)}  \!+\! \sum\limits_{v \in {\bf{V}}}^{} {\varphi _{k,s,v}^{vUAV}\left( t \right)}  \!+\! \varphi _{k,s}^{vLEO}\left( t \right) \!\le\! 1,\forall k \in {{\bf{K}}},\\
&~~~~ C4:{\bm{\rho _s}},{\bm{\eta _s}} \!\in\! \left( {0,1} \right),\!\forall s \!\in\! \{\! 1,2,3\!\} \!,\!\\
&~~~~ C5:p_{k,s,m,n}^{vBS}\left( t \right),p_{k,s,v,n}^{vUAV}\left( t \right),p_{k,s,n}^{vLEO}\left( t \right) \!\ge\! 0, \forall k \in {\bf{K}},n \in {\bf{N}},m \in {\bf{M}},v \in {\bf{V}},\\
&~~~~ C6:\xi _{k,s,n}^{}\!\left( t \right),{\varphi}_{k,s,m(v)}^{}\!\left( t \right) \!\in\! \left\{ {0,1} \right\},\forall k \!\in\! {\bf{K}},n \!\in\! {\bf{N}},m \!\in\! {\bf{M}},v \!\in\! {\bf{V}},\!\\
&~~~~ C7:P_{m,s}^{vBS}\left( t \right) \!\le\! \rho _s^{vBS}\left( t \right)P_B^{},\forall m \!\in\! {\bf{M}},  P_{v,s}^{vUAV}\left( t \right) \!\le\! \rho _s^{vUAV}\left( t \right)P_V^{},\forall v \!\in\! {\bf{V}}, P_s^{vLEO}\left( t \right) \!\le\! \rho _s^{vLEO}\left( t \right)P_L^{},\forall s \!\in\! \{ 1,2,3\} ,\!\\
&~~~~ C8: \sum\limits_{n \in {\bf{N}}}^{} {\sum\limits_{k \in {{\bf{K}}_s}}^{} {\xi _{k,s,n}^{}\left( t \right){\varphi}_{k,s,m}^{v\!B\!S}\left( t \right) \!\le\! \eta _s^{v\!B\!S}\!\left( t \right)N} } \!,\! \sum\limits_{n \in {\bf{N}}}^{} {\sum\limits_{k \in {{\bf{K}}_s}}^{} {\xi _{k,s,n}^{}\left( t \right){\varphi}_{k,s,v}^{v\!U\!A\!V}\left( t \right) \!\le\! \eta _s^{v\!U\!A\!V}\!\left( t \right)N} } \!,\! \\
&~~~~~~~~~~ \sum\limits_{n \in {\bf{N}}}^{} {\sum\limits_{k \in {{\bf{K}}_s}}^{} {\xi _{k,s,n}^{}\left( t \right){\varphi}_{k,s}^{v\!L\!E\!O}\left( t \right) \!\le\! \eta _s^{v\!L\!E\!O}\!\left( t \right)N} } \!,\! \forall m \!\in\! {\bf{M}},\! v \!\in\! {\bf{V}},\! s \!\in\! \{ 1,2,3\} \!,\!\\
&~~~~ C9:{\left( {x_i^{vUAV}(t) \!-\! x_j^{vUAV}(t)} \right)^2} \!+\! {\left( {y_i^{vUAV}(t) \!-\! y_j^{vUAV}(t)} \right)^2} \!\ge\! {\left( {d_{\min }^{vUAV}} \right)^2},\forall i,j \in {\bf{V}},i \ne j,\\
&~~~~ C10:\rho _1^{vBS(vUAV)(vLEO)}(t) + \rho _2^{vBS(vUAV)(vLEO)}(t) \!+\! \rho _3^{vBS(vUAV)(vLEO)}(t) \!\le\! 1 ,\!\\
&~~~~ C11:\eta _1^{vBS(vUAV)(vLEO)}(t) +\eta _2^{vBS(vUAV)(vLEO)}(t) \!+\! \eta _3^{vBS(vUAV)(vLEO)}(t) \!\le\! 1.\!
\end{aligned}
\end{equation}
\hrulefill
\end{figure*}

In this context, we define:
\begin{equation} \label{32b}
\setlength{\abovedisplayskip}{3pt}
\setlength{\belowdisplayskip}{3pt}
{{\bf{P}}_{\bf{s}}} \!=\! \left\{ {{\bf{P}}_{\bf{s}}^{{\bf{v}}{\bf{B}}{\bf{S}}}(t),{\bf{P}}_{\bf{s}}^{{\bf{v}}{\bf{U}}{\bf{A}}{\bf{V}}}(t),{\bf{P}}_{\bf{s}}^{{\bf{v}}{\bf{L}}{\bf{E}}{\bf{O}}}(t)} \right\}.
\end{equation}
Furthermore, in \eqref{31}, $\beta$ is the initial threshold used for guaranteeing the non-negativity of the delay objective.
In \eqref{31}, $C1$ exhibits that a subchannel is only allocated to at most one user in the same cell and slice,
$C2$ shows that we can only assign at most one subchannel to one user in the same TS,
and $C3$ represents that a user is only associated with one network component in the same TS.
Still referring to \eqref{31}, $C4$ represents the value range of the allocation factors $\{{\bm{\eta _s}}\!,{\bm{\rho_s}}\}$.
Furthermore, $C5$ limits the non-negativity of the transmit power,
while $C6$ limits the value range of binary variables $\xi_{k,s,n}\!\left( t \right)$ and ${\varphi}_{k,s,m(v)}^{}\!\left( t \right)$.
To elaborate further, $C7$ illustrates that the actual power consumed by each vLEO satellite, vUAV and vBS could not exceed the total available power on class-$s$ slices, respectively,
while $C8$ guarantees that the number of subchannels used in each network component on class-$s$ slices at TS $t$ is no more than the total subchannel numbers.
In order to avoid collision, $C9$ limits the distance between two vUAVs to no less than $d_{\min }^{vUAV}$, where we assume that all UAVs in the network are at the same altitude.
Finally, $C10$ and $C11$ illustrate that the power as well as subchannels assigned to different slice types do not overlap with each other in any of the TSs.

However, since we both consider the inter-cell interferences and introduce the binary variables $\left\{ {{\bf{\Xi }}_{\bf{s}}}\!(t)\!,\!{\bf{\Phi }}\!_{\bf{s}}\!(t) \right\}$ for our heterogeneous 3D network, \eqref{31} represents a non-convex and mixed-integer non-scalar MOOP, which is subject to multiple constraints.
It is challenging to solve it by traditional mathematical optimization methods, and it is also ineffective to directly utilize existing DRL algorithms.
\vspace{-0.05in}
\section{Problem Transformation and Solution}
In this section, firstly, we decompose \eqref{31} into two sub-problems in order to reduce its complexity.
Then, based on the MADDPG algorithm \cite{16}, we develop a CDMADDPG model and propose the CDMADDPG algorithm to solve \eqref{31} as well as to obtain near-Pareto optimal solutions.

\vspace{-0.1in}
\subsection{MOOP Decoupling}
The proposed Eq.~\eqref{31} includes 6 sets of decision variables and 3 optimization objectives, which makes the exploration of the full Pareto front excessively complex.
MOOP decoupling can help mitigate this problem \cite{3.3}.
Furthermore, the RAN slices are independent with each other, so the users' resource allocation within a slice is not affected by other slices.
Therefore, we can directly decouple \eqref{31} into the inter-slice optimization subproblem $\bf{P_1}$ and the intra-slice optimization subproblem $\bf{P_2}$, as follows:

\begin{equation}\label{34}
\setlength{\abovedisplayskip}{1.5pt}
\setlength{\belowdisplayskip}{-1pt}
\begin{aligned}
&{{\bf{P_1}}}:\!\mathop {\max }\limits_{\left\{ {\scriptstyle{\bf{d}}_{\bf{s}}^{{\bf{\!v\!U\!A\!V\!}}},{\bm{\eta _s}}\!,{\bm{\rho_s}}\hfill} \right\}}
\left\{ \!\begin{split}
&\sum\nolimits_{t =\! 1}^T {} R_1^{sum}\!\left( t \right)\\
&\sum\nolimits_{t =\! 1}^T {} \left[ {\beta  \!-\! D_2^{ave}\!\left( t \right)} \right]\\
&\sum\nolimits_{t =\! 1}^T {} \!S\!I\!N\!R_3^{ave}\!\left( t \right)
\end{split} \!\right\},\\
&\text{s.t.}
~ C4,C9\!-\!C11,
\end{aligned}
\end{equation}
and
\begin{equation}\label{35}
\setlength{\abovedisplayskip}{1.5pt}
\setlength{\belowdisplayskip}{0pt}
\begin{aligned}
&{{\bf{P_2}}}:\!\mathop {\max }\limits_{\left\{ {{{\bf{\Xi }}_{\bf{s}}}{\bf{,}}{{\bf{\Phi }}_{\bf{s}}}{\bf{,}}{{\bf{P}}_{\bf{s}}}} \right\}} \left\{ \!\begin{split}
&\sum\nolimits_{t =\! 1}^T {} R_1^{sum}\!\left( t \right),s \!=\! 1\\
&\sum\nolimits_{t =\! 1}^T {} \left[ {\beta  \!-\! D_2^{ave}\!\left( t \right)} \right],s \!=\! 2\\
&\sum\nolimits_{t =\! 1}^T {} \!S\!I\!N\!R_3^{ave}\!\left( t \right),s \!=\! 3
\end{split} \!\right\},\\
&\text{s.t.}
~ C1 \!-\! C3,C5\!-\!C8.\\
\end{aligned}
\end{equation}

Firstly, the subproblem $\bf{P_1}$ allocates the resources $\left\{ {\bf{d}}_{\bf{s}}^{{\bf{vUAV}}},{{\bm{\eta }}_{\bf{s}}},{{\bm{\rho }}_{\bf{s}}} \right\}$ among slices in order to pursue Pareto-optimization of multiple performance metrics.
Then, according to $\left\{ {\bf{d}}_{\bf{s}}^{{\bf{vUAV}}},{{\bm{\eta }}_{\bf{s}}},{{\bm{\rho }}_{\bf{s}}} \right\}$, the subproblem $\bf{P_2}$ allocates the users' resources $\left\{ {{{\bf{\Xi }}_{\bf{s}}},{{\bf{\Phi }}_{\bf{s}}},{{\bf{P}}_{\bf{s}}}} \right\}$ within the class-$s$ slices to optimize the respective performance of each slice, facilitating a beneficial performance enhancement.
In this way, $\bf{P_1}$ and $\bf{P_2}$ are optimized iteratively for gradually approaching the Pareto-optimization of the original problem $\bf{P^*}$.
When the number of iterations is large enough, the optimal solutions to the two subproblems can be regarded as optimal also for the original problem.

\vspace{-0.1in}
\subsection{The CDMADDPG Model Structure}
\begin{figure}[!tp]
\setlength{\abovecaptionskip}{0cm}
\setlength{\belowcaptionskip}{0cm}
  \centering
  \includegraphics[width=3.5in]{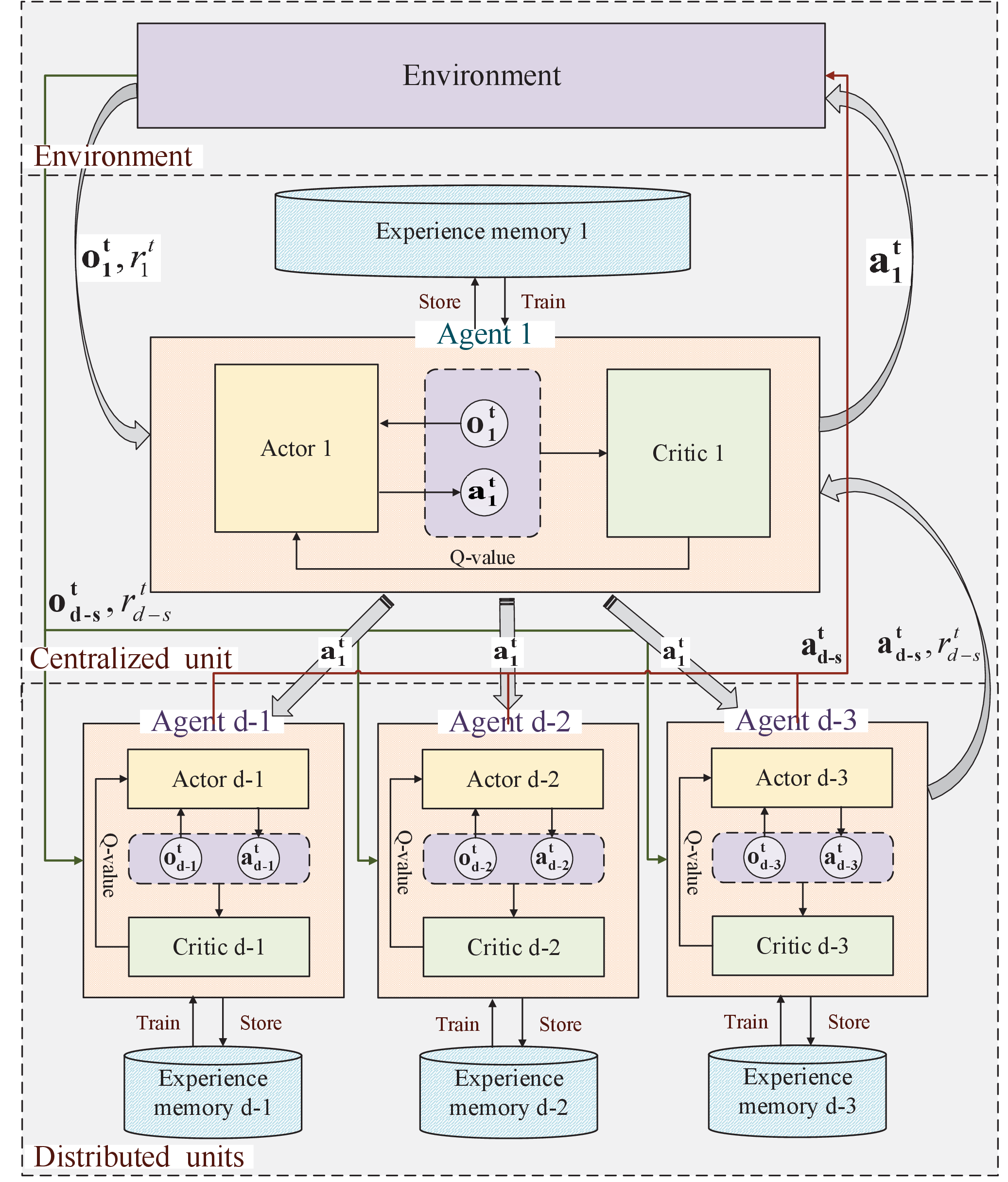}\\
  \caption{CDMADDPG model structure.}\label{fig2}
\end{figure}

In this section, the CDMADDPG model structure is proposed for solving the above decoupled subproblems, as shown in Fig. \ref{fig2}.
It is composed of three parts, namely the environment, the upper centralized unit and three underlying distributed units.
Each centralized or distributed unit relies on DRL, including a main body, termed as Agent $i,\! \forall  i \!\in\! \{1,d\!-\!1,d\!-\!2,d\!-\!3\}$, which is equivalent to a player in a game.
Agent $i$ can interact with the environment for formulating real-time state observations and reward functions, and may rely on different reward functions to facilitate MOO.
In the CDMADDPG model, it is assumed that each agent's tasks exhibit Markovian attributes.
Explicitly, a Markov decision process (MDP) consists of a four-tuple, namely $\left\langle {\bf{o}}_{\bf{i}}^{\bf{t}},{\bf{a}}_{\bf{i}}^{\bf{t}},r_{i}^t,{\bf{o}}_{\bf{i}}^{{\bf{t \!+\! 1}}} \right\rangle \!,\! \forall  i \!\in\! \{1,\!d\!-\!1,\!d\!-\!2,\!d\!-\!3\}$,
the elements of which are the set of state observation, the action space, reward functions and the next state observation, respectively.
In the MDP, the agent observes the current state of the environment and selects an action from the action space.
By executing this action, the agent receives a reward from the environment and then goes to the next state.
However, we cannot collect global action information for directly determining the optimal MDP, given the huge continuous action space of the CDMADDPG model.
Hence, the agents have to train deep neural networks (DNNs) and learn the optimal decisions through their continuous interactions with the environment.

Bearing in mind that the dynamic resource sharing among slices based upon the same physical SAGIN is similar to the multi-agent cooperation scenario of the CDMADDPG model, the global MNO in the system may be viewed as Agent 1 of the centralized supervisory unit in the CDMADDPG model to carry out the inter-slice resource optimization for the entire SAGIN.
Similarly, the three classes of SAGIN slices in the system may be transformed into Agent $d$-$s,\!s\!\in\!\{1,2,3\}$ of the underlying distributed units in the CDMADDPG model in order to process the resource allocation within a specific type of slices, respectively.

Then, the elements of the system model may also correspond to the MDP's four-tuples of $\!\left\langle\! {\bf{o}}_{\bf{i}}^{\bf{t}},{\bf{a}}_{\bf{i}}^{\bf{t}},r_{i}^t,{\bf{o}}_{\bf{i}}^{{\bf{t + 1}}}\!\right\rangle \!$.

\textbf{The current and the next state of Agent 1:} As for Agent 1, the state observation is held in a vector of the users' request and position dynamics in the entire system.
Hence, at the time step $t$ (equivalent to TS $t$), the state observation of Agent 1 is defined as:
\begin{equation}\label{36}
{\bf{o}}_{\bf{1}}^{\bf{t}} \!=\! \left\{ {{\bf{A}}\!\left( {\bf{t}} \right),\left( {{\bf{x}}_{}^{{\bf{UE}}}{\bf{(t)}},{\bf{y}}_{}^{{\bf{UE}}}{\bf{(t)}}} \right)} \right\},\!
\end{equation}
where ${\bf{A}}\!\left( {\bf{t}} \right) \!=\! \left\{ {{A_1}\!\left( t \right),{A_2}\!\left( t \right),{A_3}\!\left( t \right)} \right\}$ is the set of the data arrival process of the three types of slices in the system.
Furthermore, $\left( {{\bf{x}}_{}^{{\bf{UE}}}{\bf{(t)}},{\bf{y}}_{}^{{\bf{UE}}}{\bf{(t)}}} \right) \!=\! \left\{ {\left( {x_{k,s}^{UE}(t),y_{k,s}^{UE}(t)} \right)} \right\}$ represents the coordinate set of all the terrestrial users.

Then, the next state observation of Agent 1 can be formulated as:
\begin{equation}\label{37}
{\bf{o}}_{\bf{1}}^{{\bf{t \!+\! 1}}} \!=\! \left\{ {{\bf{A}}\!\left( {{\bf{t \!+\! 1}}} \right),\left( {{\bf{x}}_{}^{{\bf{UE}}}{\bf{(t \!+\! 1)}},{\bf{y}}_{}^{{\bf{UE}}}{\bf{(t \!+\! 1)}}} \right)} \right\},\!
\end{equation}
where ${{\bf{A}}}\!\left( {\bf{t+1}} \right)$ and $\left( {{\bf{x}}_{}^{{\bf{UE}}}{\bf{(t \!+\! 1)}},\!{\bf{y}}_{}^{{\bf{UE}}}{\bf{(t \!+\! 1)}}} \right)$ represent the data arrival process of the three types of slices and all users' coordinates at the next time step.

\textbf{Action of Agent 1:} At the time step $t$, the MNO has to allocate optimal subchannels and power resources for each class of slices, and determine the optimal vUAVs' positions according to the state observation ${\bf{o}}_{\bf{1}}^{\bf{t}}$.
Hence, the action set is denoted as:
\begin{equation}\label{38}
{\bf{a}}_{\bf{1}}^{\bf{t}} \!=\! \left\{ {{{\bm{\eta }}_{\bf{s}}},{{\bm{\rho }}_{\bf{s}}},{\bf{d}}_{\bf{s}}^{{\bf{vUAV}}}} \right\},\!\forall s \!\in\! \left\{ {1,2,3} \right\}.\!
\end{equation}
In order to further reduce the action space, we replace ``$\le$" in $C10$ and $C11$ of \eqref{34} with ``$=$".
In this way, Agent 1 only has to train the power and subchannel allocation factors of the first two kinds of slices.
Note that the selected actions have to satisfy the constraints of ${{\bf{P_1}}}$, where those unsatisfied actions are removed.

\textbf{State and the next state of Agent $\bm{d}$-$\bm{s}$:} As for Agent $d$-$s$ in the distributed units, the state observation is a vector of users' request on the class-$s$ slices.
Hence, at the time step $t$, the state observation of Agent $d$-$s$ is defined as:
\begin{equation}\label{39}
{\bf{o}}_{{\bf{d \!-\! s}}}^{\bf{t}} \!=\! \left\{ {A_{1,s}\!\left( t \right)\!,\!A_{2,s}\!\left( t \right)\!,\!...,\!A_{k,s}\!\left( t \right)\!,\!...} \right\},
\end{equation}
where ${A_{k,s}}\left( t \right)$ is the random data arrival of user $k$ on class-$s$ slices at time step $t$.

Naturally, the next state observation of Agent $d$-$s$ can be formulated as:
\begin{equation}\label{40}
{\bf{o}}_{{\bf{d \!-\! s}}}^{{\bf{t \!+\! 1}}} \!=\! \left\{ {A_{1,s}\!\left( t\!+\!1 \right)\!,\!A_{2,s}\!\left( t\!+\!1 \right)\!,\!...,\!A_{k,s}\!\left( t\!+\!1 \right)\!,\!...} \right\}.
\end{equation}

\textbf{Action of Agent $\bm{d}$-$\bm{s}$:} At the time step $t$, the class-$s$ slices are mapped to the suitable vBSs/vUAVs/vLEO satellite and rely on optimal subchannels and power resources for their users within the slices according to the state observation ${\bf{o}}_{\bf{d\!-\!s}}^{\bf{t}}$. Hence, the action set is denoted as:
\begin{equation}\label{41}
{\bf{a}}_{{\bf{d \!-\! s}}}^{\bf{t}} \!=\! \left\{ {{{\bf{\Xi }}_{\bf{s}}},{{\bf{\Phi }}_{\bf{s}}},{{\bf{P}}_{\bf{s}}}} \right\}.\!
\end{equation}
For facilitating our mathematical analysis, we relax binary variables of ${{\bf{\Xi }}_{\bf{s}}}$ and ${{\bf{\Phi }}_{\bf{s}}}$ to continuous variables within the value range of [0,1], which also conforms to the condition of continuous actions in our CDMADDPG algorithm.
Similarly, the selected actions have to satisfy the constraints of ${{\bf{P_2}}}$.
Since sometimes resource contention occurs, a dual resource allocation mechanism is proposed for improving the algorithm's performance and the users' QoS.
Specifically, within a time step $t$, the first resource allocation of Agent $d$-$s$ is carried out to obtain ${\bf{a}}_{{\bf{d \!-\! s}}}^{\bf{t}}$.
Then, the algorithm checks whether the action set satisfies the constraints, and collects all idle resources.
If some of the actions violate the resource constraints and there are resources available in the system, a second resource allocation round shall be carried out immediately to resolve these conflicting actions in order to avoid that the related users fail to get the requested service due to collisions.
However, if all available resources in the system are used up, then except for one of the conflicting actions, which retains its original value, all conflicting actions ${{\bf{\Xi }}_{\bf{s}}}$ and ${{\bf{\Phi }}_{\bf{s}}}$ are set to 0 and ${{{\bf{P}}_{\bf{s}}}}$ are changed to a low value, in order to satisfy the constraints.

\textbf{Reward of Agent $\bm{d}$-$\bm{s}$:} Since the main performance metrics of the three classes of RAN slices are different, we define different reward functions for Agent $d$-$s$ in the CDMADDPG model.
According to \eqref{35}, the reward space of Agent $d$-$s$ can be formulated as follows:
\begin{equation} \label{42}
r_{d-s}^t \!=\! \left\{ \begin{split}
&\overline{R_1^{sum}}\left( t \right),s \!=\! 1,\\
&\left[ {1  \!-\! \overline{D_2^{ave}}\left( t \right)} \right],s \!=\! 2,\\
&\overline{S\!I\!N\!R_3^{ave}}\left( t \right),s \!=\! 3,
\end{split} \right.
\end{equation}
where $\overline{R_1^{sum}}\!\left( t \right)$, $\overline{D_2^{ave}}\!\left( t \right)$ and $\overline{S\!I\!N\!R_3^{ave}}\!\left( t \right)$ are the normalized objectives by 0-1 normalization.
Then, the experience memory of the Agent $d$-$s$ is summarized as ${{{\bf{D}}_{\bf{d-s}}}}$, the elements of which are $\left\langle\! {\bf{o}}_{\bf{d-s}}^{\bf{t}},{\bf{a}}_{\bf{d-s}}^{\bf{t}},r_{d-s}^t,{\bf{o}}_{\bf{d-s}}^{{\bf{t + 1}}}\!\right\rangle $.

\textbf{Reward of Agent 1:} According to \eqref{34}, Agent 1 aims for finding Pareto optimal solutions of our MOOP.
Hence, we define the reward function for Agent 1 by harnessing the ensemble strategy of rank voting method \cite{34}.
Specifically, we first define the experience memory of Agent 1 as ${{{\bf{D}}_{\bf{1}}}}$ to store four-tuples $\! \left\langle \! {{\bf{o}}_{\bf{1}}^{\bf{t}},\!{\bf{a}}_{\bf{i}}^{\bf{t}},\!r_{i}^t,\!{\bf{o}}_{\bf{1}}^{{\bf{t \!+\! 1}}}} \! \right\rangle$ having $ i \!=\! 1,\!d\!-\!1,\!d\!-\!2,\!d\!-\!3$ generated by the entire model.
Then, we rank the four-tuples in ${\bf{D}}_{\bf{1}}$ in ascending order according to the reward value $r_{d \!-\! s}^t,\!\forall s \!\in\! \left\{ {\!1,2,3\!} \right\}$, respectively.
Hence, the index ${\bf{\chi }}_{{s}}^t$ is assigned to each four-tuple, which means the four-tuple $\left\langle {{\bf{o}}_{\bf{1}}^{\bf{t}},\!{\bf{a}}_{\bf{i}}^{\bf{t}},\!r_{i}^t,\!{\bf{o}}_{\bf{1}}^{{\bf{t + 1}}}} \right\rangle$ is ranked ${\bf{\chi }}_{{s}}^t$-th position in the ascending sequence by $r_{d - s}^t$, as follows:
\begin{equation} \label{43}
\begin{split}
&\chi _{{s}}^t \!=\! \textbf{asc} \left( {\bf{D}}_{\bf{1}}; \left\langle {{\bf{o}}_{\bf{1}}^{\bf{t}},\!{\bf{a}}_{\bf{i}}^{\bf{t}},\!r_{i}^t,\!{\bf{o}}_{\bf{1}}^{{\bf{t + 1}}}} \right\rangle \right) \!,\!\\
&\textrm{where} \, i \!=\! 1,\!d\!-\!1,\!d\!-\!2,\!d\!-\!3 .\!
\end{split}
\end{equation}
In \eqref{43}, $\textbf{asc} \left( \textbf{X}; y \right)$ is the function ordering the $\textbf{X}$'s elements in ascending order,
and then feeding back the ascending index of the element $y$ to $\chi _{{s}}^t$, where $y$ is one of the elements in $\textbf{X}$.
By adding ${\bf{\chi }}_{{s}}^t$ of three classes of slices for each time step, we can obtain the ${\bf{\chi }}_{}^t$ as the reward function $r_1^t$, as follows:
\begin{equation} \label{44}
r_1^t \!=\!\chi _{}^t \!=\! \chi _1^t \!+\! \chi _2^t \!+\! \chi _3^t.
\end{equation}
In this way, we keep approaching the optimal four-tuple, as follows:
\begin{equation} \label{45b}
\left\langle \! {{\bf{o}}_{\bf{1}}^{*},\!{\bf{a}}_{\bf{i}}^{*},\!r_{i}^{*},\!{\bf{o'}}_{\bf{1}}^{*}} \! \right\rangle  \!=\! \mathop {\arg \max }\limits_t  {\bf{\chi }}_{}^{t},\!
\end{equation}
where $\left\langle \! {{\bf{o}}_{\bf{1}}^{*},\!{\bf{a}}_{\bf{i}}^{*},\!r_{i}^{*},\!{\bf{o'}}_{\bf{1}}^{*}} \! \right\rangle$ is not dominated in the space of ${\bf{D}}_{\bf{1}}$, which conforms to the definition of Pareto optimal solutions (Please refer to Appendix A for the specific proof).

\textbf{Basic structure of centralized and distributed units:}
In our CDMADDPG model, both the centralized and the three distributed DRL units rely on the same basic actor-critic (AC) structure \cite{16}, where each agent contains an actor and a critic, as seen from Fig. \ref{fig2}.
Specifically, the actor selects an action by the current state observation; then the critic aims to evaluate this action and return the Q-value; and finally the actor modifies subsequent action selection policies based on the critic's Q-value.
In this way, the learning process of the CDMADDPG model becomes more stable and converges faster than the DRL models operating without the AC structure.
Among them, the actor $i,\! \forall  i \!\in\! \{1,\!d\!-\!1,\!d\!-\!2,\!d\!-\!3\}$ is designated as the policy function ${\mu _{{\theta _i}}}\!\left( {{\bf{o}}_{\bf{i}}^{\bf{t}}|{\theta ^{{\mu _i}}}} \right)$ of the DRL unit $i$, while the critic $i$ could be defined as the action-value function ${Q_{{\theta _i}}}\!\left( {{\bf{o}}_{\bf{i}}^{\bf{t}},{\bf{a}}_{\bf{i}}^{\bf{t}}|{\theta ^{{Q_i}}}} \right)$ of the DRL unit $i$.
Thus, we utilize different DNNs for approximating the actor $i$ and the critic $i$, respectively, and use the classic stochastic gradient method to train both ${\theta ^{{\mu _i}}}$ and ${\theta ^{{Q_i}}}$.
Actor $i$'s input is set to the current state ${\bf{o}}_{\bf{i}}^{\bf{t}}$ and its output is set to the deterministic action ${\bf{a}}_{\bf{i}}^{\bf{t}}$;
while critic $i$'s inputs are the state ${\bf{o}}_{\bf{i}}^{\bf{t}}$ and the action ${\bf{a}}_{\bf{i}}^{\bf{t}}$ generated by the actor, and its output is set to the estimated $Q$-value.
Hence, we could obtain the optimal reward $r_i^t$ by continuously updating both the actor and the critic of Agent $i$.

\textbf{Updating the actor:}
The OF of agent $i$ in the CDMADDPG algorithm can be defined as the expectation of long-term discounted cumulative reward, that is:
\begin{equation} \label{46}
{J_i}\left( \mu  \right) = {E_\mu }\left[ {r_i^0 + \gamma r_i^1 + {\gamma ^2}r_i^2 +  \cdots  + {\gamma ^{T \!-\! 1}}r_i^{T \!-\! 1}} \right],
\end{equation}
where $\gamma$ is defined as the discount factor.
The actor $i$ aims to find the optimal deterministic policy $\mu^*_{{\theta _i}}$, which corresponds to the maximization of ${J_i}\left( \mu \right)$, formulated as:
\begin{equation} \label{47}
{\mu ^*_{{\theta _i}}}\left( {{\bf{o}}_{\bf{i}}^{\bf{t}}|{\theta ^{{\mu _i}}}} \right) = \mathop {\arg \max }\limits_\mu  {J_i}\left( \mu  \right).
\end{equation}
Hence, the deterministic action ${\bf{a}}_{\bf{i}}^{\bf{t}}$ of agent $i$ at time step $t$ could be obtained through ${\mu _{{\theta _i}}^*}$, as follows:
\begin{equation} \label{48}
{\bf{a}}_{\bf{i}}^{\bf{t}} = {\mu _{{\theta _i}}^*}\left( {{\bf{o}}_{\bf{i}}^{\bf{t}}|{\theta ^{{\mu _i}}}} \right).
\end{equation}

It has been proved in \cite{35} that the gradient of ${J_i }\left( \mu  \right)$ with respect to ${\theta ^{{\mu _i}}}$ is equivalent to the expected gradient of ${Q_{{\theta _i}}}\!\left( {{\bf{o}}_{\bf{i}}^{\bf{t}},{\bf{a}}_{\bf{i}}^{\bf{t}}|{\theta ^{{Q_i}}}} \right)$ with respect to ${\theta ^{{\mu _i}}}$.

As for the action-value function ${Q_{{\theta _i}}}\!\left( {{\bf{o}}_{\bf{i}}^{\bf{t}},{\bf{a}}_{\bf{i}}^{\bf{t}}|{\theta ^{{Q_i}}}} \right)$ of agent $i$, based on MDP, it can be formally defined using the Bellman's equation, as follows:
\begin{equation} \label{49}
{Q_{{\theta _i}}}\!\left( {{\bf{o}}_{\bf{i}}^{\bf{t}},{\bf{a}}_{\bf{i}}^{\bf{t}}|{\theta ^{{Q_i}}}} \right) \!=\! {r_i^t} \!+\! \gamma \mathop {\max }\limits_{\bf{a}} {Q_{{\theta _i}}}\!\left( {{\bf{o}}_{\bf{i}}^{\bf{t\!+\!1}},{\bf{a}}_{\bf{i}}^{\bf{t\!+\!1}}|{\theta ^{{Q_i}}}} \right)\!,\!
\end{equation}
where ${r_i^t}$ is the current reward function,
while ${\bf{o}}_{\bf{i}}^{\bf{t+1}}$ and ${\bf{a}}_{\bf{i}}^{\bf{t+1}}$ are the next state and the next action of agent $i$, respectively.

Subsequently, the update of ${Q_{{\theta _i}}}\!\left( {{\bf{o}}_{\bf{i}}^{\bf{t}},{\bf{a}}_{\bf{i}}^{\bf{t}}|{\theta ^{{Q_i}}}} \right)$ is as follows:
\begin{equation} \label{50}
{Q_{{\theta _i}}}\!\left( {{\bf{o}}_{\bf{i}}^{\bf{t}},{\bf{a}}_{\bf{i}}^{\bf{t}}|{\theta ^{{Q_i}}}} \right) \!\leftarrow\! {Q_{{\theta _i}}}\!\left( {{\bf{o}}_{\bf{i}}^{\bf{t}},{\bf{a}}_{\bf{i}}^{\bf{t}}|{\theta ^{{Q_i}}}} \right) \!+\! \alpha {\delta _t},
\end{equation}
where
\begin{equation} \label{51}
{\delta _t} \!=\! r_i^t \!+\!\gamma \mathop {\max }\limits_{\bf{a}} {Q'_{{\theta _i}}}\!\left( {{\bf{o}}_{\bf{i}}^{\bf{t\!+\!1}},{\bf{a}}_{\bf{i}}^{\bf{t\!+\!1}}|{\theta ^{{Q'_i}}}} \right) \!-\! {Q_{{\theta _i}}}\!\left( {{\bf{o}}_{\bf{i}}^{\bf{t}},{\bf{a}}_{\bf{i}}^{\bf{t}}|{\theta ^{{Q_i}}}} \right)\!,\!
\end{equation}
and $r_i^t \!+\! \gamma \mathop {\max }\limits_{\bf{a}} {Q'_{{\theta _i}}}\!\left( {{\bf{o}}_{\bf{i}}^{\bf{t+1}},{\bf{a}}_{\bf{i}}^{\bf{t+1}}|{\theta ^{{Q'_i}}}} \right)$ is defined as the target, which represents the actual reward of the prediction.
Furthermore, ${\delta _t}$ is the error, which is used for estimating the action-value function, while $\alpha$ is the learning rate.

Therefore, the chain rule could be utilized to derive ${J_i}\!\left( \mu  \right)$ and then to get the update method of the actor, as follows:
\begin{equation} \label{52}
\begin{split}
&{\nabla _{{\theta ^{{\mu _i}}}}}J_i = {E_{{\bf{o}}_{\bf{i}}^{\bf{t}},{\bf{a}}_{\bf{i}}^{\bf{t}}\sim{{\bf{D}}_{\bf{i}}}}}\left[ {\nabla _{{\theta ^{{\mu _i}}}}}{\mu _{{\theta _i}}}\left( {{\bf{o}}_{\bf{i}}^{\bf{t}}|{\theta ^{{\mu _i}}}} \right)\right.\\
&\left.\cdot{\nabla _{{\bf{a}}_{\bf{i}}^{\bf{t}}}}{Q_{{\theta _i}}}\left( {{\bf{o}}_{\bf{i}}^{\bf{t}},{\bf{a}}_{\bf{i}}^{\bf{t}}|{\theta ^{{Q_i}}}} \right){|_{{\bf{a}}_{\bf{i}}^{\bf{t}} = {\mu _{{\theta _i}}}\left( {{\bf{o}}_{\bf{i}}^{\bf{t}}|{\theta ^{{\mu _i}}}} \right)}} \right].
\end{split}
\end{equation}

Referring to the gradient ascent algorithm, we could update the parameter ${\theta ^{{\mu _i}}}$ in \eqref{52} and optimize the action ${\bf{a}}_{\bf{i}}^{\bf{t}}$ along the direction of improving the action-value function ${Q_{{\theta _i}}}\left( {{\bf{o}}_{\bf{i}}^{\bf{t}},{\bf{a}}_{\bf{i}}^{\bf{t}}|{\theta ^{{Q_i}}}} \right)$.

\textbf{Updating the critic:}
According to \eqref{50} and \eqref{51}, we update the critic by minimizing the loss function, defined as follows:
\begin{equation} \label{53}
{L_i} \!=\! {E_{{\bf{o}}_{\bf{i}}^{\bf{t}},{\bf{a}}_{\bf{i}}^{\bf{t}},r_i^t,{\bf{o}}_{\bf{i}}^{{\bf{t}} \!+\! 1}\!\sim\!{{\bf{D}}_{\bf{i}}}}}\left[ {{{\left( {{\rm{Target}}\,{Q_i} \!-\! {Q_{{\theta _i}}}\!\left( {{\bf{o}}_{\bf{i}}^{\bf{t}},\!{\bf{a}}_{\bf{i}}^{\bf{t}}|{\theta ^{{Q_i}}}} \right)} \right)}^2}} \right]\!,\!
\end{equation}
and
\begin{equation} \label{54}
{\rm{Target}}\,{Q_i} = r_i^t + \gamma Q_{{\theta _i}}'\left( {{\bf{o}}_{\bf{i}}^{{\bf{t + 1}}},\mu _{{\theta _i}}'\left( {{\bf{o}}_{\bf{i}}^{{\bf{t + 1}}}|{\theta ^{\mu _i'}}} \right)|{\theta ^{Q_i'}}} \right),
\end{equation}
where ${\theta ^{\mu _i'}}$ and ${\theta ^{Q_i'}}$ in ${\rm{Target}}\,{Q_i}$ are the parameters from the target actor and the target critic, respectively.

Therefore, the gradient of the critic can be expressed as shown below:
\begin{equation} \label{55}
\begin{split}
&{\nabla_{{\theta ^{{Q_i}}}}}L_i \!=\! {E_{{\bf{o}}_{\bf{i}}^{\bf{t}}\!,{\bf{a}}_{\bf{i}}^{\bf{t}}\!,r_i^t\!,{\bf{o}}_{\bf{i}}^{{\bf{t}}\! +\! 1}\!\sim\!{{\bf{D}}_{\bf{i}}}}}\!\left[ \left( {\rm{Target}}\,{Q_i} - \right.\right. \\
&\left.\left. {Q_{{\theta _i}}}\!\left( {{\bf{o}}_{\bf{i}}^{\bf{t}},\!{\bf{a}}_{\bf{i}}^{\bf{t}}|{\theta ^{{Q_i}}}}\! \right) \right)\!{\nabla _{{\theta ^{{Q_i}}}}}\!{Q_{{\theta _i}}}\!\left( {{\bf{o}}_{\bf{i}}^{\bf{t}},{\bf{a}}_{\bf{i}}^{\bf{t}}|{\theta ^{{Q_i}}}} \right)\! \right].\!
\end{split}
\end{equation}

The gradient descent method is used to update the parameter ${\theta ^{{Q_i}}}$ and then to find the optimal value ${Q^*_{{\theta _i}}}$.

\begin{table*}[!ht]
\scriptsize
\centering
\caption*{Table 3: DNNs' configuration in this paper.}
\begin{tabular}{|c|c|c|c|c|c|c|c|c|}
  \hline
   & Number of inputs & Number of outputs & \makecell[c]{Number of \\hidden layers} & \makecell[c]{Number of \\neurons} & Activation function & Learning rate & \makecell[c]{Fixed size \\of $\bf{D_i}$} & $D_i^{mini}$\\
  \hline
  Actor 1 & $(3\!+\!6K_s)$ & $(12\!+\!2V)$ & 2 & 100 & Relu+Sigmoid & 0.0001 & \multirow{2}{*}{10,000}& \multirow{2}{*}{100}\\
  \cline{1-7}
  Critic 1 &$(15\!+\!6K_s\!+\!2V)$ & 1&2 & 100 & Relu+Sigmoid & 0.001 & &\\
  \hline
  Actor $d$-$s$ & $K_s$&$3K_s$ &2 & 100 & Relu+Sigmoid & 0.0001 & \multirow{2}{*}{2000}& \multirow{2}{*}{50}\\
  \cline{1-7}
  Critic $d$-$s$ & $4K_s$& 1&2 & 100 & Relu+Sigmoid & 0.001 & &\\
  \hline
\end{tabular}
\end{table*}
\begin{table*}[!ht]
\scriptsize
\centering
\caption*{Table 4: Specific values of the parameters in \eqref{58b}.}
\begin{tabular}{|c|c|c|c|c|c|c|c|c|c|c|}
  \hline
  Parameter & $N_{ag}$ & ${L_{i,ac}}$ & ${L_{i,cri}}$ & $n_{1,ac}^{\left( 0 \right)}$ & $n_{1,ac}^{\left( 1 \right)}$& $n_{1,ac}^{\left( 2 \right)}$&  $n_{1,ac}^{\left( 3 \right)}$& $n_{1,cri}^{\left( 0 \right)}$&$n_{1,cri}^{\left( 1 \right)}$& $n_{1,cri}^{\left( 2 \right)}$ \\
   \hline
  Value & 4& 2 & 2 & $(3\!+\!6K_s)$& 100& 100&  $(12\!+\!2V)$ & \makecell[c]{$(15\!+\!6K_s$\\$\!+2V)$}&100& 100\\
   \hline
  Parameter & $n_{1,cri}^{\left( 3 \right)}$& $n_{d\!-\!s,ac}^{\left( 0 \right)}$ & $n_{d\!-\!s,ac}^{\left( 1 \right)}$& $n_{d\!-\!s,ac}^{\left( 2 \right)}$&  $n_{d\!-\!s,ac}^{\left( 3 \right)}$& $n_{d\!-\!s,cri}^{\left( 0 \right)}$ &  $n_{d\!-\!s,cri}^{\left( 1 \right)}$&  $n_{d\!-\!s,cri}^{\left( 2 \right)}$& $n_{d\!-\!s,cri}^{\left( 3 \right)}$& \\
  \hline
  Value & 1& $K_s$& 100& 100&  $3K_s$&  $4K_s$& 100 &  100& 1& \\
  \hline
\end{tabular}
\end{table*}

\textbf{Online network and target network:}
In order to guarantee the stability of the learning process, the CDMADDPG algorithm creates two DNNs for each actor/critic, \textit{i.e.} the online and target networks, as follows:
\begin{equation} \label{56}
{\rm{actor}}\left\{ \begin{split}
&{\rm{online}}:{\mu _{{\theta _i}}}\left( {{\bf{o}}_{\bf{i}}^{\bf{t}}|{\theta ^{{\mu _i}}}} \right),{\rm{update}}\;{\theta ^{{\mu _i}}},\\
&{\rm{target}}:\mu _{{\theta _i}}'\left( {{\bf{o}}_{\bf{i}}^{\bf{t}}|{\theta ^{\mu _i'}}} \right),{\rm{update}}\;{\theta ^{\mu _i'}},
\end{split} \right.
\end{equation}
and
\begin{equation} \label{57}
{\rm{critic}}\left\{ \begin{split}
&{\rm{online}}:{Q_{{\theta _i}}}\left( {{\bf{o}}_{\bf{i}}^{\bf{t}},{\bf{a}}_{\bf{i}}^{\bf{t}}|{\theta ^{{Q_i}}}} \right),{\rm{update}}\;{\theta ^{{Q_i}}},\\
&{\rm{target}}:Q_{{\theta _i}}'\left( {{\bf{o}}_{\bf{i}}^{\bf{t}},{\bf{a}}_{\bf{i}}^{\bf{t}}|{\theta ^{Q_i'}}} \right),{\rm{update}}\;{\theta ^{Q_i'}}.
\end{split} \right.
\end{equation}
More explicitly, the CDMADDPG algorithm uses a total of 16 DNNs ($(1[\textrm{online network}]+ 1[\textrm{target network}])\times 2[\textrm{actor\&critic}]\times4[\textrm{agents}]$).
Firstly, the agent $i$ updates ${\theta ^{\mu _i}}$ and ${\theta ^{Q_i}}$ of the online networks through the gradient ascent/descent method after finishing a mini-batch based data training.
Then, ${\theta ^{\mu _i'}}$ and ${\theta ^{Q_i'}}$ of the target networks are updated using the following soft update method:
\begin{equation} \label{58}
\left\{ \begin{split}
&{\theta ^{Q_i'}} \leftarrow \tau {\theta ^{{Q_i}}} + \left( {1 - \tau } \right){\theta ^{Q_i'}},\\
&{\theta ^{\mu _i'}} \leftarrow \tau {\theta ^{{\mu _i}}} + \left( {1 - \tau } \right){\theta ^{\mu _i'}},
\end{split} \right.
\end{equation}
where the soft update parameter $\tau$ is 0.001.

\textbf{DNN configuration:}
According to the defined state observation and action space of Agent $i$, we set different number of DNNs' inputs and outputs for respectively approximating Actors as well as Critics.
Then, the fixed size of the experience memory is limited to $\bf{{D_1}}$$ =\! 10,000$ four-tuples and $\bf{{D_{d-s}}}$$ =\! 2000$ four-tuples, and we sample $D_1^{mini}\!=\!100$ mini-batches as well as $D_{d-s}^{mini} \!=\! 50$ mini-batches for training the centralized unit and distributed units at each time step, respectively.
However, no known rule is used for determining the number of neurons and hidden layers, hence it is chosen by trial-and-error \cite{36}.
For instance, we set Relu activation function for the hidden layer and then set Sigmoid activation function for the output layer, because all the outputs are positive and between 0 and 1.
The specific DNNs' configuration is summarized in Table 3.

\vspace{-0.1in}
\subsection{The CDMADDPG Algorithm}
Based on the above CDMADDPG model, the specific algorithmic process is formulated for approaching the Pareto optimal solutions of our MOOP.
The CDMADDPG algorithm is executed by the centralized supervisory unit for global learning between different types of slices, and uses the slice-specific distributed units to make local resource decisions within the slices, respectively.

Firstly, Agent 1 in the centralized unit selects the action ${\bf{a}}_{\bf{1}}^{\bf{t}}$ based on the current policy and exploration for inter-slice resource optimization, and delivers it to Agent $d$-$s$.
In the distributed units, the three dedicated slice-specific agents decide their own actions ${\bf{a}}_{\bf{d-s}}^{\bf{t}}$ for intra-slice resource optimization, and obtain the current reward $r_{d-s}^{t}$ and the next state ${\bf{o}}_{d-s}^{{\bf{t + 1}}}$ by their interaction with the environment in a parallel and asynchronous manner.
The four-tuples of Agent $d$-$s$ generated are stored in ${{{\bf{D}}_{\bf{d-s}}}}$.
We randomly sample $D_{d-s}^{mini}$ mini-batches selected from ${{{\bf{D}}_{\bf{d-s}}}}$ to critic $d$-$s$ and actor $d$-$s$ for online and off-policy training.
After delivering ${\bf{a}}_{\bf{d-s}}^{\bf{t}}$ and $r_{d-s}^{t}$ of Agent $d$-$s$ to Agent 1 and collecting the next state ${\bf{o}}_{1}^{{\bf{t + 1}}}$ from the environment, all the four-tuples of the system are generated and stored in ${\bf{D_1}}$.
Next, we randomly sample $D_{1}^{mini}$ mini-batches selected from ${\bf{D_1}}$ and find the non-dominated four-tuple, \textit{i.e.}, calculating $r_1^t$ by \eqref{44}, for training the critic and actor of Agent 1.
In this way, we alternately train the centralized unit and the distributed units until convergence is reached.
Finally, it is hard to offer any proof of evidence that the Pareto optimal solutions have indeed been found \cite{36a,36c}.
Hence we resort to collecting all the non-dominated four-tuples $\left\langle \! {{\bf{o}}_{\bf{1}}^{*},\!{\bf{a}}_{\bf{i}}^{*},\!r_{i}^{*},\!{\bf{o'}}_{\bf{1}}^{*}} \! \right\rangle$ of a few of the final iterations as the near-Pareto optimal solutions of the MOOP formulated.
We summarize the specific process in Algorithm 1.
By activating Algorithm 1 several times, we can collect multiple near-Pareto optimal solutions so as to present an approximation of the Pareto boundary.

\vspace{0.1in}
\textbf{\textit{Theorem 1:}}
Similar to the investigations in \cite{36b}, the order of computational complexity of the learning procedure in our proposed algorithm is given by
\begin{equation} \label{58b}
\mathcal{O}\left( {ET\sum\limits_{i = 1}^{{N_{ag}}} {\left( {\sum\limits_{j = 0}^{{L_{i,ac}}} {n_{i,ac}^{\left( j \right)}n_{i,ac}^{\left( {j + 1} \right)}}  + \sum\limits_{j = 0}^{{L_{i,cri}}} {n_{i,cri}^{\left( j \right)}n_{i,cri}^{\left( {j + 1} \right)}} } \right)} } \right),
\end{equation}
where $E$ and $T$ refer to the number of episodes and time steps, respectively.
Furthermore, $N_{ag}$ represents the number of Agents in our CDMADDPG algorithm, while $n_{i,ac}^{\left( j \right)}$ and $n_{i,cri}^{\left( j \right)}$ are defined as the numbers of neurons in the $j$-th layer of actor $i$ and critic $i$, respectively.
Still referring to \eqref{58b}, ${L_{i,ac}}$ (${L_{i,cri}}$) represents the number of hidden layers in actor (critic) $i$.
\vspace{0.1in}

In our algorithm, the specific values of these parameters are
shown in Table~4, which correspond to the DNNs' configuration in Table~3.
We substitute the values of Table~4 into \eqref{58b} to finally obtain the computational complexity order of the CDMADDPG algorithm as ${\cal O}\left({\left( {9{K_s} \!+\! V} \right)\!\times \!400 ET} \right)$.

Furthermore, the centralized Reward $r_1^t$ is a monotonically increasing function of the distributed Reward $r_{d-s}^t$, as shown in \eqref{44} and \eqref{42}, respectively. Therefore, the relationship between them can be characterized as a mixing network, as follows:
\begin{equation} \label{59}
r_1^t = {f_{mix}}\left( {r_{d - 1}^t,r_{d - 2}^t,r_{d - 3}^t} \right),
\end{equation}
where $f_{mix}$ denotes the Mixing network \cite{42,43}, which is a non-negative non-linear monotonic mapping between $r_1^t$ and $r_{d - s}^t, s \!\in\! \{1,2,3\}$.
The convergence of this centralized and distributed algorithm to a locally optimal policy has been substantiated in \cite{43,44}.
In the next section, we will continue our analysis by combining the computational complexity with the simulation of convergence.

\begin{algorithm}[!tp]
\caption{The Specific Process of our CDMADDPG Algorithm solving \eqref{31}.} \label{alg:1}
\begin{algorithmic}[1]
\STATE \textbf{Initialization:}
\STATE ~~~Randomly initialize the parameters ${\theta ^{{Q_i}}}$ and ${\theta ^{{\mu _i}}}$ of critic ${Q_{{\theta _i}}}\left( {{\bf{o}}_{\bf{i}}^{\bf{t}},{\bf{a}}_{\bf{i}}^{\bf{t}}|{\theta ^{{Q_i}}}} \right)$ and of actor ${\mu _{{\theta _i}}}\left( {{\bf{o}}_{\bf{i}}^{\bf{t}}|{\theta ^{{\mu _i}}}} \right)$ for Agent $i$.
\STATE ~~~Initialize the target network $Q_{{\theta _i}}'$ as well as $\mu _{{\theta _i}}'$, and set their parameters as ${\theta ^{Q_i'}} \leftarrow  {\theta ^{{Q_i}}}$ and ${\theta ^{\mu _i'}} \leftarrow {\theta ^{{\mu_i}}}$.
\STATE ~~~Initialize the experience memory ${{{\bf{D}}_{\bf{i}}}}$ of Agent $i$.
\STATE \textbf{For} episode = 1, ..., $E$ \textbf{do}:
\STATE ~~~Initialize a random process ${{\bf{N}}^{\bf{t}}_{\bf{i}}}$ for action exploration.
\STATE ~~~Obtain the initial set of states ${\bf{o}}_{i}^{\bf{0}}$.
\STATE ~~~\textbf{For} time step = 0, ..., ${T \!-\! 1}$ \textbf{do}:
\STATE ~~~~~~As for Agent 1, select action ${\bf{a}}_{\bf{1}}^{\bf{t}} = {\mu _{{\theta _{1}}}}\left( {{\bf{o}}_{\bf{1}}^{\bf{t}}|{\theta ^{{\mu_{1}}}}} \right) + {{\bf{N}}^{\bf{t}}_{\bf{1}}}$ based on the current policy and exploration for inter-slice resource optimization.
\STATE ~~~~~~Transfer the action ${\bf{a}}_{\bf{1}}^{\bf{t}}$ of Agent 1 to Agent $d$-$s$.
\STATE ~~~~~~As for Agent $d$-$s$, select action ${\bf{a}}_{\bf{d-s}}^{\bf{t}} = {\mu _{{\theta _{d-s}}}}\left( {{\bf{o}}_{\bf{d-s}}^{\bf{t}}|{\theta ^{{\mu_{d\!-\!s}}}}} \right) + {{\bf{N}}^{\bf{t}}_{\bf{d-s}}}$ based on the current policy, on exploration and Agent 1's action ${\bf{a}}_{\bf{1}}^{\bf{t}}$ for intra-slice resource optimization.
\STATE ~~~~~~Estimate whether ${\bf{a}}_{\bf{d-s}}^{\bf{t}}$ meets the constraints of ${{\bf{P_2}}}$, otherwise activate the dual resource allocation mechanism.
\STATE ~~~~~~Perform actions ${\bf{a}}_{\bf{d-s}}^{\bf{t}}$ and record the rewards $r_{d-s}^{t}$ as well as the new states ${\bf{o}}_{\bf{d-s}}^{{\bf{t + 1}}}$.
\STATE ~~~~~~In the experience memory ${{{\bf{D}}_{\bf{d-s}}}}$, store the four-tuples of $\left\langle {\bf{o}}_{\bf{d-s}}^{\bf{t}},{\bf{a}}_{\bf{d-s}}^{\bf{t}},r_{d-s}^t,{\bf{o}}_{\bf{d-s}}^{\bf{t + 1}} \right\rangle $.
\STATE ~~~~~~Randomly sample $D_{d-s}^{mini}$ mini-batches of $\left\langle {\bf{o}}_{\bf{d-s}}^{\bf{j}},{\bf{a}}_{\bf{d-s}}^{\bf{j}},r_{d-s}^j,{\bf{o}}_{\bf{d-s}}^{{\bf{j + 1}}}\right\rangle$ selected from ${{{\bf{D}}_{\bf{d-s}}}}$.
\STATE ~~~~~~Calculate ${\rm{Target}}\,{Q_{d-s}}$ by \eqref{54}.
\STATE ~~~~~~Update critic $d$-$s$ by minimizing loss function of \eqref{55}.
\STATE ~~~~~~Update actor $d$-$s$ using the gradient policy algorithm of \eqref{52}.
\STATE ~~~~~~Update parameters of the target network for Agent $d$-$s$ by \eqref{58}.
\STATE ~~~~~~Transfer the action ${\bf{a}}_{\bf{d-s}}^{\bf{t}}$ and the reward $r_{d-s}^{t}$ of Agent $d$-$s$ to Agent 1, and record new states ${\bf{o}}_{1}^{{\bf{t + 1}}}$ of Agent 1.
\STATE ~~~~~~ In the experience memory ${\bf{D_1}}$, store the four-tuples of $\! \left\langle \! {{\bf{o}}_{\bf{1}}^{\bf{t}},\!{\bf{a}}_{\bf{i}}^{\bf{t}},\!r_{i}^t,\!{\bf{o}}_{\bf{1}}^{{\bf{t \!+\! 1}}}} \! \right\rangle \!$ having $ i \!=\! 1,\!d\!-\!1,\!d\!-\!2,\!d\!-\!3$.
\STATE ~~~~~~Randomly sample $D_{1}^{mini}$ mini-batches from ${\bf{D_1}}$.
\STATE ~~~~~~Calculate $r_1^t$ by \eqref{44} and ${\rm{Target}}\,{Q_{1}}$ by \eqref{54}.
\STATE ~~~~~~Update critic 1 by minimizing loss function of \eqref{55}.
\STATE ~~~~~~Update actor 1 using the gradient policy algorithm of \eqref{52}.
\STATE ~~~~~~Update the parameters of the target network for Agent 1 by \eqref{58}.
\STATE ~~~~~~${\bf{o}}_{i}^{\bf{t}} \leftarrow  {\bf{o}}_{i}^{\bf{t+1}}$.
\STATE ~~~\textbf{End for}
\STATE ~~~\textbf{When} episode=$(E\!-\!2)$ or $(E\!-\!1)$ or $E$ \textbf{do}:
\STATE ~~~Collect all the non-dominated four-tuples according to the value of rewards $r_{d-s}^{t},\!\forall s \!\in\! \left\{ {\!1,2,3\!} \right\}$ from ${{{\bf{D}}_{\bf{1}}}}$.
\STATE \textbf{End for}
\STATE Store the final training parameters ${\theta ^{{\mu _i}}}$, ${\theta ^{{Q_i}}}$, ${\theta ^{{\mu _i'}}}$ and ${\theta ^{Q_i'}}$ of DNNs.
\end{algorithmic}
\end{algorithm}

\section{Simulation Results and Discussions}
In this section, we exhibit simulation results to verify the theoretical analysis and compare the performance attained to benchmarkers.
We adopt a 3D Euclidean coordinate system model and the terrestrial area is given by 3000 $\! \times \!$ 3000 $(m^2)$, where $K$ users are randomly distributed.
We set $M\!=\!2$ and $V\!=\!3$, where the coordinates of the terrestrial vBSs are fixed as (1000, 1000, 0)(m) and (2000, 2000, 0)(m), respectively, and the altitudes of the vUAVs and of the vLEO satellite are fixed as $z_v^{vUAV}\!(t)\!=\!100$m and $z^{vLEO}\!(t)\!=\!200,000$m, respectively.
Due to the high altitude of satellite, the distance may be approximated as $d_{k,s}^{vLEO}(t) \approx \sqrt {{{\left( {z_{}^{vLEO}\!(t)} \right)}^2}}$.
Furthermore, the maximum rate of the vLEO satellite is 100 Mbps.
The available bandwidth of different layers is $B\!=\!30$MHz, and it is divided into $N\!=\!7$ subchannels, respectively.
Unless otherwise specified, we use 20,000 iterations of $E\!=\!20$ episodes multiplying $T\!=\!1000$ time steps.
The other parameters are summarized as follows:
$P_B\!=\!10$dBW, $P_V\!=\!20$dBW, $P_L\!=\!30$dBW, $N_0\!=\!-130$dBm/Hz, $\delta\!=\!0.1$s, $\beta\!=\!0.2$s, $\alpha\!=\!1.5$, $R\!=\!6$, $h_0\!=\!-30$dB, ${d_{\min }^{vUAV}}\!=\!100$m,
$\lambda_2\!=\!(10\!\times\!K_2)$kbps.

In order to highlight the advantages of our proposed scheme, we include four benchmark schemes for comparison.
\begin{itemize}
  \item Firstly, Benchmark 1 is the conventional single-layer and agent-coupled ``MADDPG" algorithm relying on three parallel distributed Agents, where each Agent includes an Actor, which only needs local information for its policy decisions, and a Critic, which has to collect all agents' global policies for action value learning.
  \item Secondly, in Benchmark 2, we consider the traditional scalar method of solving our MOOP. From the perspective of MNO, we convert the key metrics on three classes of slices into the network-wide utility through the weighted sum method, defined as:
\begin{equation} \label{59}
U\left( t \right) \!=\! {\omega _1}R_1^{sum}\left( t \right) \!-\! {\omega _2}D_2^{ave}\left( t \right) \!+\! {\omega _3}S\!I\!N\!R_3^{ave}\left( t \right),\!
\end{equation}
where $\omega_1$, $\omega_2$ and $\omega_3$ are the unit prices of throughput, delay and SINR gains charged by the MNO, respectively.
Clearly, the influence of $R_1^{sum}\left( t \right)$, $D_2^{ave}\left( t \right)$ and $S\!I\!N\!R_3^{ave}\left( t \right)$ is directly dependent on the weights of $\omega_1$, $\omega_2$ and $\omega_3$, respectively, which are usually defined through experience.
Therefore, we adjust their weight ratio into 1:1:1, 1:1:4 and 4:4:1, termed as ``utility(1:1:1)", ``utility(1:1:4)" and ``utility(4:4:1)" schemes, respectively, and then we use the deep deterministic policy gradient (DDPG) algorithm to solve them.
  \item Thirdly, we consider the ``single resource allocation mechanism" to be Benchmark 3 of the proposed dual resource allocation mechanism.
In Benchmark 3, when some variables in a given action set have resource collision, only one of these variables retains its original value, while the other conflicting variables will be set either to 0 or to a low value to meet the constraints, which means that the corresponding users will not be served.
  \item Finally, in Benchmark 4, we utilize the ``fixed vUAVs' position based scheme", in which the vUAVs' positions are fixed as (500, 500), (1500, 1500) and (2500, 2500), respectively;
       while in our scheme, the vUAVs' positions are dynamically optimized in $\bf{P^*}$.
\end{itemize}
We also summarize the differences of the application of these schemes in Table~5.

\begin{table*}[!ht]
\scriptsize
\centering
\caption*{Table 5: Comparison between the proposed scheme and four benchmark schemes.}
\begin{tabular}{|l|c|c|c|c|c|}
  \hline
   & Proposed scheme &\tabincell{c}{Benchmark 1 \cite{16}}& \tabincell{c}{Benchmark 2 \cite{14}} & \tabincell{c}{Benchmark 3 \cite{49}} &
   \tabincell{c}{Benchmark 4 \cite{6}}\\
   \hline
  \tabincell{l}{Optimization \\objective} & non-scalar ${{\bf{P}}^*}$ & non-scalar ${{\bf{P}}^*}$ & scalar \eqref{59} & non-scalar ${{\bf{P}}^*}$ & non-scalar ${{\bf{P}}^*}$ \\
   \hline
  \tabincell{l}{Optimization \\algorithm} & CDMADDPG &MADDPG & DDPG & CDMADDPG & CDMADDPG\\
   \hline
  \tabincell{l}{Algorithm \\characteristics} & \tabincell{c}{Hierarchical; \\4 Agents independent} &\tabincell{c}{Single layer;\\ 3 Agents coupled} & One Agent& \tabincell{c}{Hierarchical; \\4 Agents independent} & \tabincell{c}{Hierarchical; \\4 Agents independent}  \\
  \hline
  Actor's input & $\left\langle {\bf{o}}_{\bf{i}}^{\bf{t}}\right\rangle$ & $\left\langle {\bf{o}}_{\bf{s}}^{\bf{t}}\right\rangle$ & $\left\langle {\bf{o}}^{\bf{t}}\right\rangle$& $\left\langle {\bf{o}}_{\bf{i}}^{\bf{t}}\right\rangle$&  $\left\langle {\bf{o}}_{\bf{i}}^{\bf{t}}\right\rangle$ \\
  \hline
  Actor's output & $\left\langle {\bf{a}}_{\bf{i}}^{\bf{t}} \right\rangle$ & $\left\langle {\bf{a}}_{\bf{s}}^{\bf{t}} \right\rangle$ & $\left\langle {\bf{a}}^{\bf{t}} \right\rangle$ & $\left\langle {\bf{a}}_{\bf{i}}^{\bf{t}} \right\rangle$ & $\left\langle {\bf{a}}_{\bf{i}}^{\bf{t}} \right\rangle$ \\
  \hline
  Critic's input & $\left\langle {\bf{o}}_{\bf{i}}^{\bf{t}},{\bf{a}}_{\bf{i}}^{\bf{t}}\right\rangle$ & $\left\langle {\bf{o}}_{\bf{s}}^{\bf{t}},{\bf{a}}_{\bf{s}}^{\bf{t}}, {\bf{a}}_{{\bf{S\backslash s}}}^{\bf{t}} \right\rangle$ & $\left\langle {\bf{o}}^{\bf{t}},{\bf{a}}^{\bf{t}}\right\rangle$ & $\left\langle {\bf{o}}_{\bf{i}}^{\bf{t}},{\bf{a}}_{\bf{i}}^{\bf{t}}\right\rangle$ &  $\left\langle {\bf{o}}_{\bf{i}}^{\bf{t}},{\bf{a}}_{\bf{i}}^{\bf{t}}\right\rangle$\\
  \hline
  Critic's output & Q-value of Agent $i$ & Q-value of Agent $s$ & Q-value&Q-value of Agent $i$ & Q-value of Agent $i$ \\
  \hline
  \tabincell{l}{Resource allocation \\mechanism} & \tabincell{c}{Dual resource \\allocation } & \tabincell{c}{Dual resource \\allocation }& \tabincell{c}{Dual resource \\allocation }& \tabincell{c}{Single resource \\allocation }&  \tabincell{c}{Dual resource \\allocation }\\
  \hline
  vUAVs' position & \tabincell{c}{Dynamic \\optimization in ${{\bf{P}}^*}$}& \tabincell{c}{Dynamic \\optimization in ${{\bf{P}}^*}$}& \tabincell{c}{Dynamic \\optimization in ${{\bf{P}}^*}$}& \tabincell{c}{Dynamic \\optimization in ${{\bf{P}}^*}$}& \tabincell{c}{Fixed as \\(500, 500), \\(1500, 1500) and \\(2500, 2500)} \\
  \hline
  \tabincell{l}{Markers in \\Figures}& ``CDMADDPG"& ``MADDPG"& \tabincell{c}{``Utility(1:1:1)"\\``Utility(1:1:4)"\\``Utility(4:4:1)"}& \tabincell{c}{``Single resource \\allocation mechanism"}& \tabincell{c}{``Fixed vUAVs \\position scheme"} \\
  \hline
\end{tabular}
\end{table*}

\begin{figure}[!tp]
\setlength{\abovecaptionskip}{0cm}
\setlength{\belowcaptionskip}{-0.5cm}
  \centering
  \includegraphics[width=2.7in]{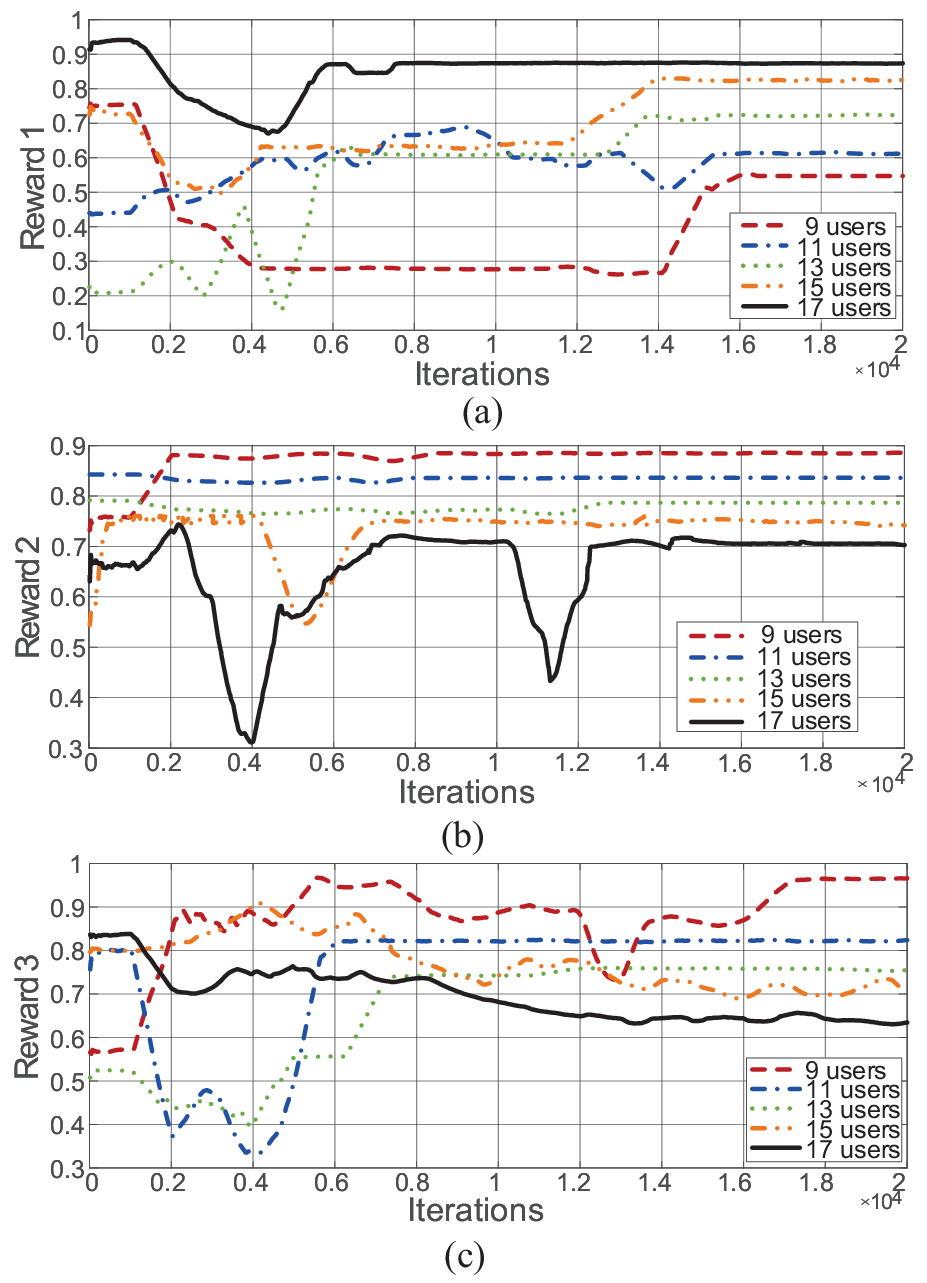}\\
  \caption{Convergence of the CDMADDPG algorithm.}\label{fig3}
\end{figure}

Fig.~\ref{fig3} shows the convergence of the CDMADDPG algorithm in the training process.
Among them, Rewards 1, 2 and 3 are three reward functions defined in \eqref{42}, which also correspond to three normalized metrics in \eqref{31}.
A total of 20,000 iterations including 20 episodes and 1000 time steps is used.
After about 15,000 training iterations, Reward 1, 2 and 3 converge to a relatively steady state, respectively.
As for the proposed non-scalar MOOP, the three conflicting optimization objectives sometimes cannot be simultaneously optimized during the iterative training process. Once resource contention occurs, there may be situations where some rewards converge to their minimum in exchange for improving either one or both of the rewards.
Additionally, the number of users $K_s$ on each class of slices also affects the reward values.
As shown in Fig.~4, Reward 1 increases while Reward 2 and 3 decrease with the increase of $K_s$ from 9 users to 17 users, which corresponds to the increase of both the class-1 slices' throughput and the class-2 slices' average delay, as well as to the reduction of the class-3 slices' average SINR, respectively.

\begin{figure}[!tp]
\setlength{\abovecaptionskip}{0cm}
\setlength{\belowcaptionskip}{-0.5cm}
  \centering
  \includegraphics[width=2.7in]{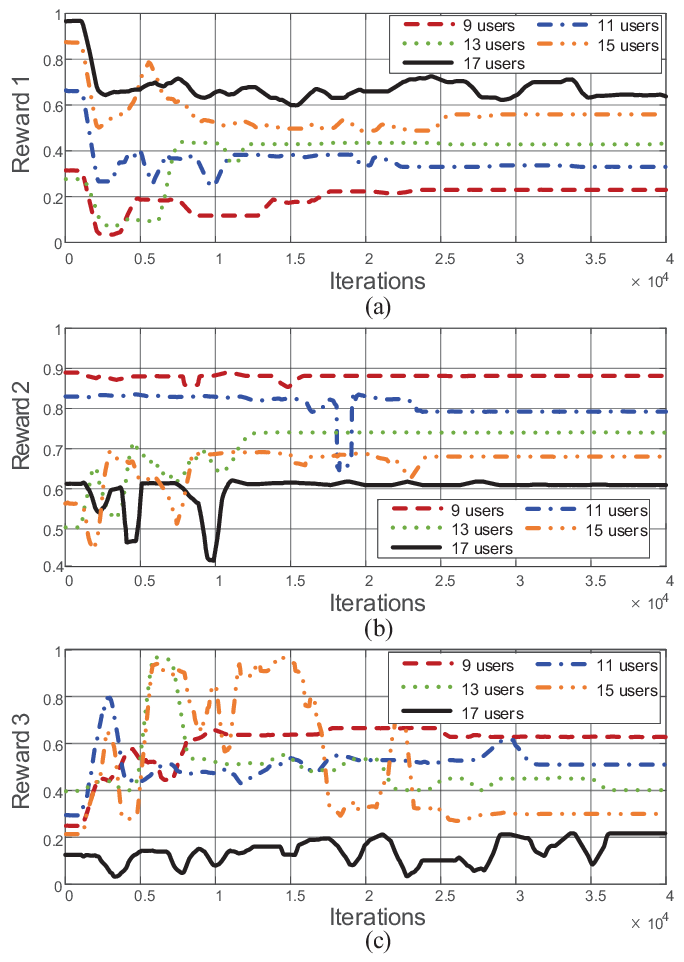}\\
  \caption{Convergence of the MADDPG algorithm.}\label{fig4}
\end{figure}

Fig.~\ref{fig4} characterizes the convergence performance of the MADDPG algorithm under the same DNN hyperparameter configuration as our algorithm.
We use a total of 40,000 iterations.
Observe that all the Rewards 1, 2 and 3 can finally converge to a relatively stable state after 36,000 training iterations.
Due to the single-layer and the agent-coupled structure of the MADDPG algorithm, each agent has to collect both local and global information at the same time.
This means that upon increasing $K_s$, the sets of both the global state observations and of the action space will rapidly expand.
This imposes a substantial computational load on the training process of the algorithm, hence making the training convergence a challenge.
We also arrive at this conclusion from Theorem 1.
According to \eqref{58b}, the computational complexity order of the CDMADDPG algorithm is given by ${\cal O}\left({\left( {9{K_s} \!+\! V} \right)\!\times \!400 ET} \right)$ for the current parameter setting, while that of the MADDPG algorithm is ${\cal O}\left({\left( {9{K_s} \!+\! 4V} \right)\!\times \!600 ET} \right)$.
We can see that the complexity difference between the proposed and MADDPG algorithm is expanding with the increase of $K_s$ and iterations $ET$.
Additionally, increasing $K_s$ has the same influence on the performance of class-$s$ slices as the proposed algorithm.

\begin{figure}[!tp]
\setlength{\abovecaptionskip}{0cm}
\setlength{\belowcaptionskip}{-0.5cm}
  \centering
  \includegraphics[width=3in]{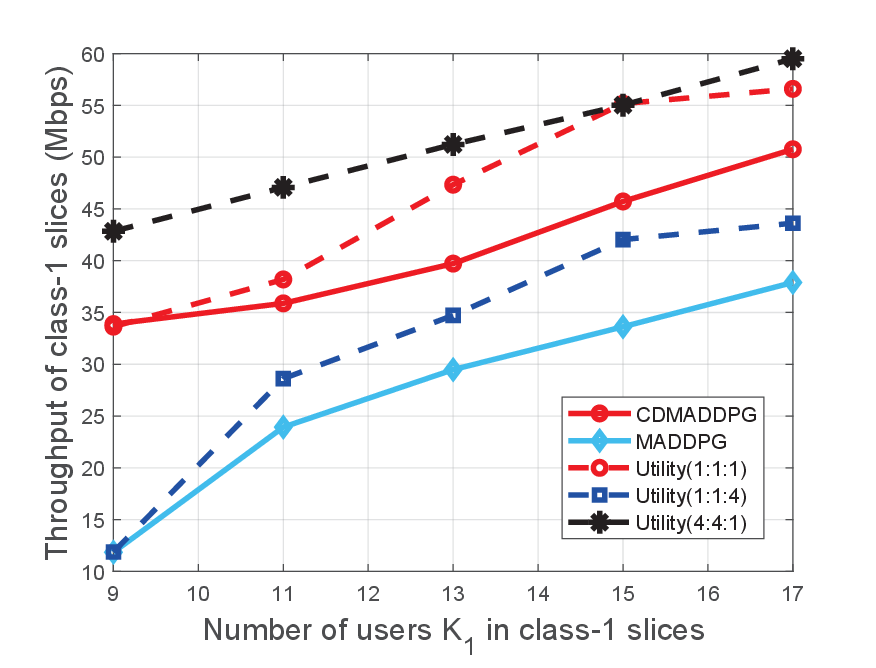}\\
  \caption{Throughput versus the number of users $K_1$ of class-1 slices.}\label{fig5}
\end{figure}

\begin{figure}[!tp]
\setlength{\abovecaptionskip}{0cm}
\setlength{\belowcaptionskip}{-0.5cm}
  \centering
  \includegraphics[width=3in]{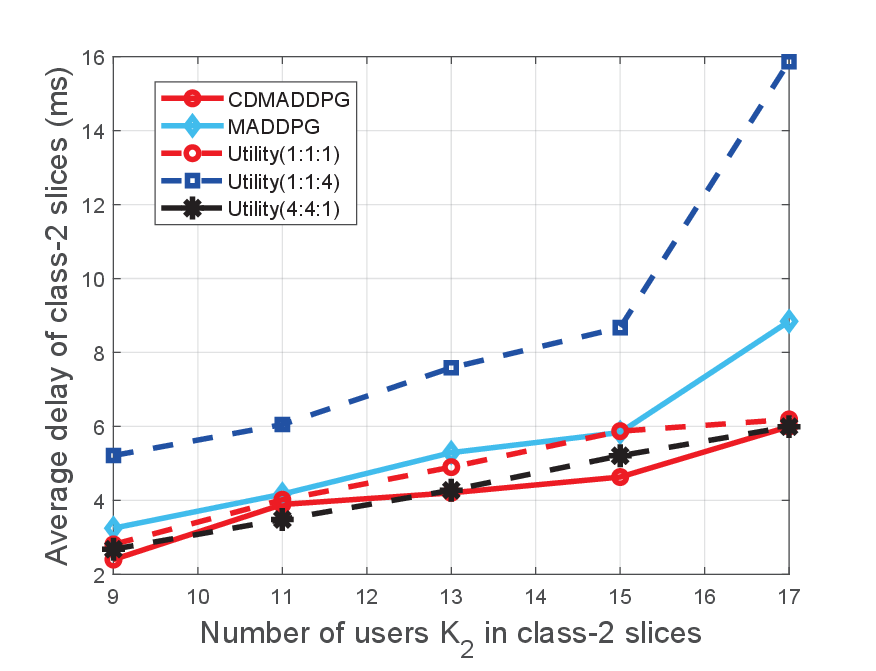}\\
  \caption{Average delay versus the number of users $K_2$ of class-2 slices.}\label{fig6}
\end{figure}

\begin{figure}[!tp]
\setlength{\abovecaptionskip}{0cm}
\setlength{\belowcaptionskip}{-0.5cm}
  \centering
  \includegraphics[width=3in]{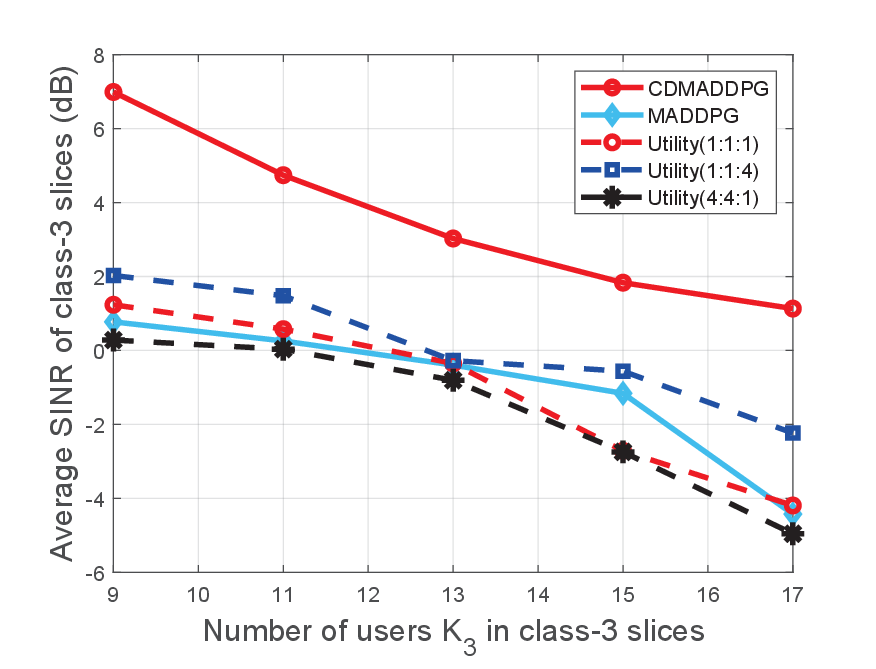}\\
  \caption{Average SINR versus the number of users $K_3$ of class-3 slices.}\label{fig7}
\end{figure}

Fig.~\ref{fig5}, Fig.~\ref{fig6} and Fig.~\ref{fig7} characterize the throughput, the average delay and average SINR versus $K_1$, $K_2$ and $K_3$ for the proposed scheme and for the benchmarkers, respectively.
In order to facilitate our comparisons, we show time-averaged values in this section, \textit{i.e.}, $\frac{1}{T}\sum\nolimits_{t =\! 1}^T {} R_1^{sum}\!\left( t \right)$, $\frac{1}{T}\sum\nolimits_{t =\! 1}^T {} Q_2^{ave}\!\left( t \right)$ and $\frac{1}{T}\sum\nolimits_{t =\! 1}^T {} \!S\!I\!N\!R_3^{ave}\!\left( t \right) $, and give the mean of 10 experiment trials with the same hyperparameter configuration.
Observe in Fig.~\ref{fig5}, Fig.~\ref{fig6} and Fig.~\ref{fig7} that with the increase of $K_s$, both the throughput and the average delay of all schemes continue to increase, whilst the average SINR is decreasing.
Compared to the MADDPG algorithm, our CDMADDPG algorithm has better performance at 20,000 iterations, thanks to its hierarchical and decoupled framework.

Observe for the three utility schemes of Fig.~\ref{fig5}, Fig.~\ref{fig6} and Fig.~\ref{fig7} that the larger the weight assigned, the better the performance becomes.
Therefore, the utility(4:4:1) scheme attains a better throughput and has the lowest average delay, but its average SINR is poor.
By contrast, the SINR advantage of the utility(1:1:4) scheme is more obvious, but as a penalty, both the throughput and the delay become inferior.
Therefore, the influence of the metrics in the utility scheme is highly dependent on their weights, which are subjectively designed in advance.
However, it is worth noting that sometimes even if we increase the weight of a specific metric, we may not witness any striking advantage.
For example, although the SINR of the utility(1:1:4) scheme is somewhat higher than that of the utility(1:1:1) scheme, its throughput and delay is gravely deteriorated.
That is to say, even though the traditional scalar method has the benefit of a priori knowledge, it still struggles to satisfy the users' QoS.

By contrast, our scheme is eminently suitable for different types of RAN slices even without designing the utility function and weights in advance.
It finds numerous near-Pareto optimal solutions, and these solutions are non-dominated by each other.
Taking into account all the above discussions regarding Fig.~\ref{fig5}, Fig.~\ref{fig6} and Fig.~\ref{fig7}, the CDMADDPG scheme is capable of striking the best tradeoff (pursuing Pareto optimal) among the three classes of slices considered, even if it does not always achieve the best result for all scenarios.


\begin{figure}[!tp]
\setlength{\abovecaptionskip}{0cm}
\setlength{\belowcaptionskip}{-0.5cm}
  \centering
  \includegraphics[width=3in]{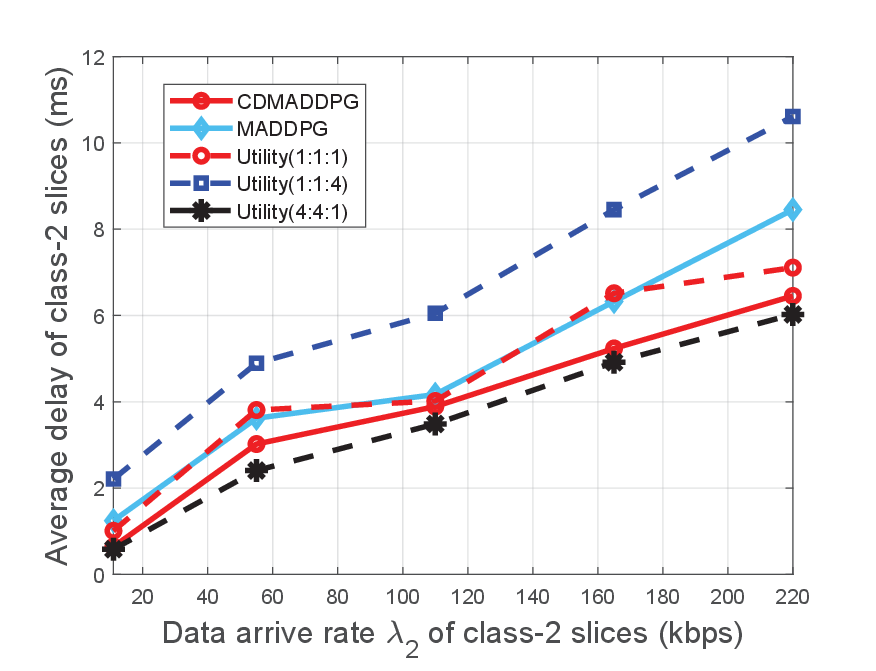}\\
  \caption{Average delay versus the data arrival rate $\lambda_2$ of class-2 slices.}\label{fig8}
\end{figure}

Fig.~\ref{fig8} portrays the average delay versus the data arrival rate $\lambda_2$ of class-2 slices for all schemes when $K_2\!=\!11$.
Upon increasing $\lambda_2$, the average delay of each scheme continues to increase.
We can always attain a lower average delay by the CDMADDPG algorithm than that of the MADDPG algorithm.

\begin{figure}[!tp]
\setlength{\abovecaptionskip}{0cm}
\setlength{\belowcaptionskip}{-0.5cm}
  \centering
  \includegraphics[width=3in]{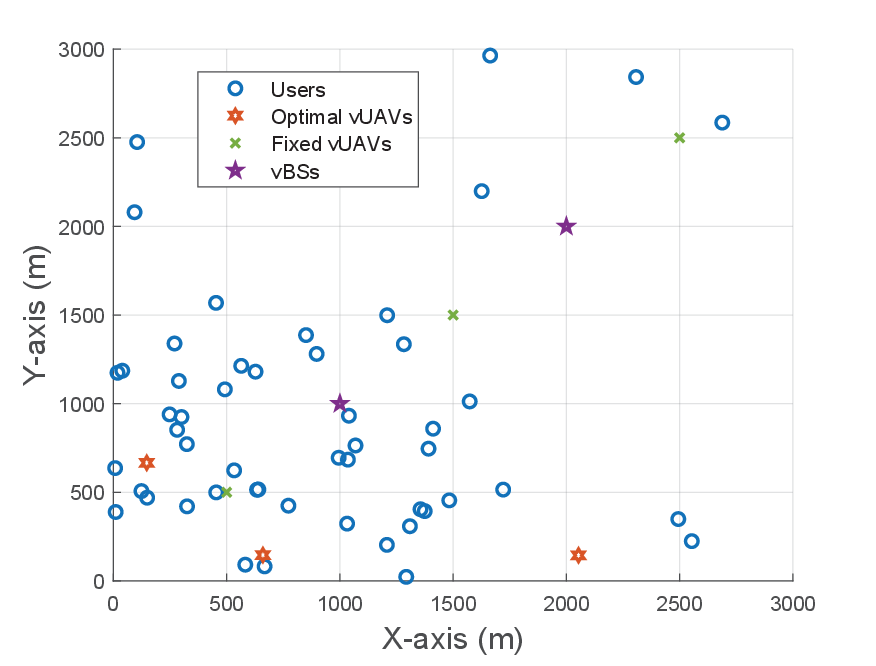}\\
  \caption{Coordinate distribution of users, optimal vUAVs, fixed vUAVs and vBSs ($K_1\!=\!K_2\!=\!K_3\!=$17).}\label{fig9}
\end{figure}

Fig.~\ref{fig9} compares the two-dimensional coordinates of vUAVs in our proposed scheme to that in the fixed vUAVs' position scheme.
In order to facilitate the analysis, we define the area by the coordinates (0, 0) to (1500, 1500)m as an urban area with a higher user density and define an area between (1500, 1,500) and (3000, 3000)m as a rural area having a small number of users, as shown in Fig.~\ref{fig9}.
Naturally, the optimal vUAVs will actively approach the high-density areas to reduce the transmission distance, but the fixed vUAVs' positions cannot be adjusted in real time.
By jointly considering the position distribution of vBSs and optimal vUAVs in Fig.~\ref{fig9}, our scheme is seen to provide a reasonable network component deployment.

\begin{figure}[!tp]
\setlength{\abovecaptionskip}{0cm}
\setlength{\belowcaptionskip}{-0.5cm}
  \centering
  \includegraphics[width=3in]{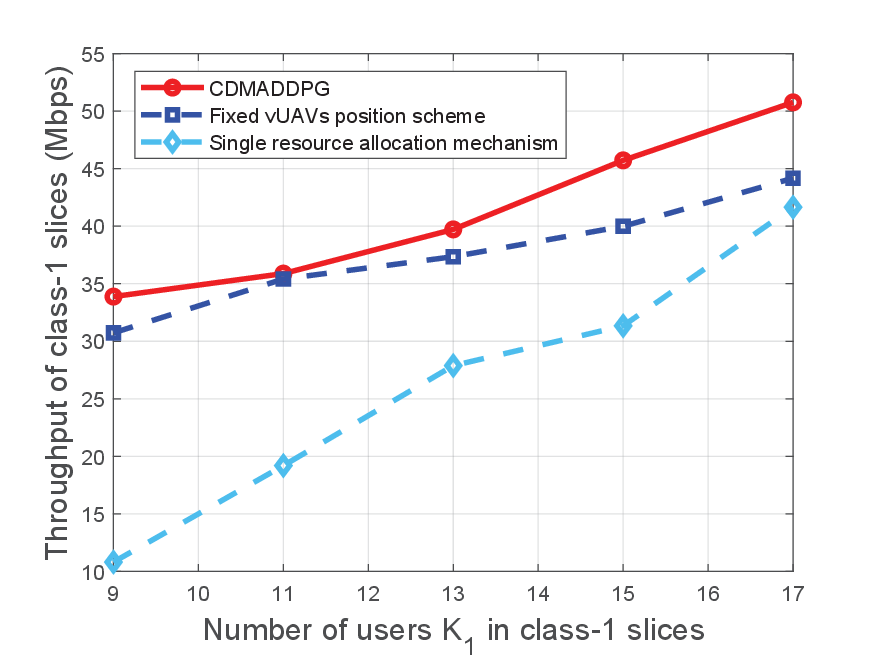}\\
  \caption{Throughput comparison with the benchmarkers.}\label{fig10}
\end{figure}

\begin{figure}[!tp]
\setlength{\abovecaptionskip}{0cm}
\setlength{\belowcaptionskip}{-0.5cm}
  \centering
  \includegraphics[width=3in]{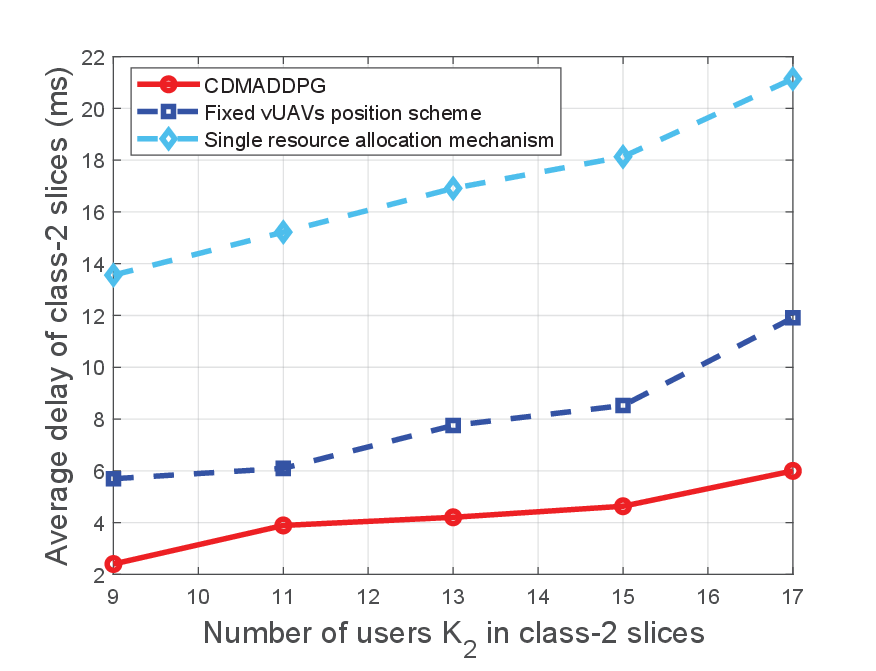}\\
  \caption{Average delay comparison with the benchmarkers.}\label{fig11}
\end{figure}

\begin{figure}[!tp]
\setlength{\abovecaptionskip}{0cm}
\setlength{\belowcaptionskip}{-0.5cm}
  \centering
  \includegraphics[width=3in]{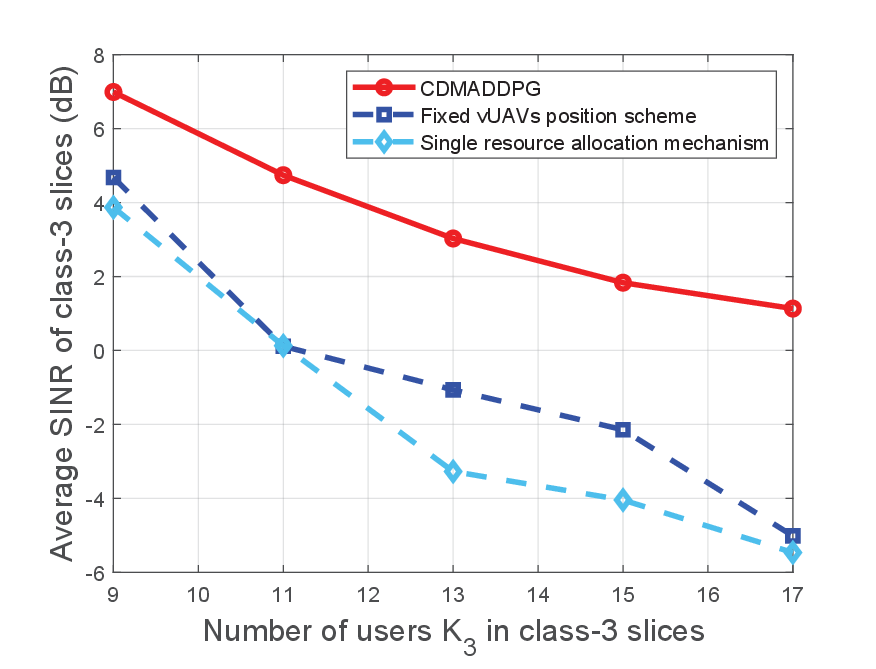}\\
  \caption{Average SINR comparison with the benchmarkers.}\label{fig12}
\end{figure}

Fig.~\ref{fig10}, Fig.~\ref{fig11} and Fig.~\ref{fig12} portray our performance comparison of the CDMADDPG scheme both with the fixed vUAVs' position based scheme and with the single resource allocation mechanism.
Firstly, it can be seen that all the performance metrics of the proposed scheme are better than those of the fixed vUAVs' position based scheme.
Explicitly, the throughput difference between both schemes is relatively small, while their average delay as well as average SINR are quite different.
It transpires that the communication distance has a grave impact on both the average delay and on the average SINR.
As expected, since we apply a dual resource allocation mechanism, the proposed scheme exhibits a higher performance advantage than the single resource allocation mechanism.
During the DRL training, a large action space leads to a high collision probability, which has a grave impact both on the algorithm's performance and on the users' QoS.
In this case, having a dual resource allocation mechanism is indispensable.

In heterogeneous SAGINs, terrestrial, aerial and satellite communications have their own service advantages, which is also reflected in our simulation results.
We collect the value of inter-slice sub-channel allocation factors ${\eta _s^{vBS}}(t)$, ${\eta _s^{vUAV}}(t)$ and ${\eta _s^{vLEO}}(t)$ when Algorithm 1 converges, as shown in Table~6.
It can be seen that class-1 slices tend to provide access by both vBSs and vUAVs; class-2 slices have a large proportion of vUAVs channel resources; finally, vLEO channel resources are mainly allocated to class-3 slices. This is in line with the service characteristics of these three communication methods.
Therefore, the proposed model is eminently suitable for our 3D network having three distinct space-air-ground channel features, and our CDMADDPG algorithm is eminently suitable for the customized services of users on the different SAGIN slices.

\begin{table}[!tp]
\centering
\caption*{Table 6: Average value of ${\eta _s^{vBS}}(t)$, ${\eta _s^{vUAV}}(t)$ and ${\eta _s^{vLEO}}(t)$ of ten simulations.}
\begin{tabular}{|c|c|c|c|}
  \hline
   & ${\eta _s^{vBS}}(t)$ & ${\eta _s^{vUAV}}(t)$ & ${\eta _s^{vLEO}}(t)$ \\
   \hline
  Class-1 slices & 0.428 & 0.333 & 0.286 \\
  \hline
  Class-2 slices & 0.286 & 0.445 & 0.143 \\
  \hline
  Class-3 slices & 0.286 & 0.222 & 0.571 \\
  \hline
\end{tabular}
\end{table}

\begin{figure}[!tp]
\setlength{\abovecaptionskip}{0cm}
\setlength{\belowcaptionskip}{-0.5cm}
  \centering
  \includegraphics[width=3.2in]{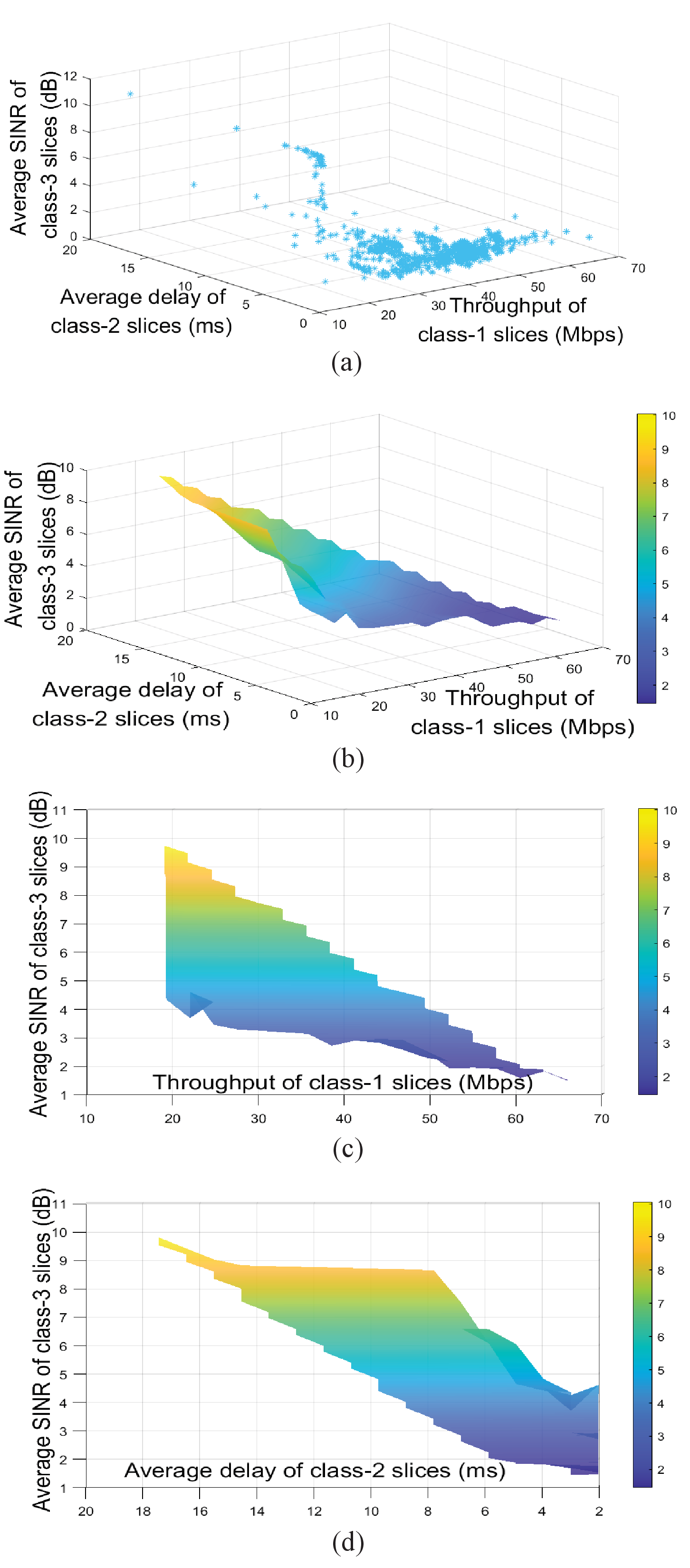}\\
  \caption{(a) Near-Pareto optimal scatter diagram; (b) Interpolated near-Pareto boundary; (c) Front view of (b); (d) Side view of (b) ($K_1\!=\!K_2\!=\!K_3\!=$11).}\label{fig13}
\end{figure}

Fig.~\ref{fig13}(a) portrays the three-dimensional scatter diagram of near-Pareto optimal solutions for \eqref{31}, which is obtained by iteratively activating Algorithm 1 ten times when $K_s\!=\!11$.
Let us briefly consider a pair of extreme points at the top left and bottom right corners represented by the approximate throughputs of [18,68]Mbps, delays of [18,2]ms and SINR of [10,1]dB.
Furthermore, Fig.~\ref{fig13}(b) represents the approximate Pareto boundary derived by using interpolation among the points of Fig.~\ref{fig13}(a).
For the convenience of observation, we also provide the front view and side view of Fig.~\ref{fig13}(b), as shown in Fig.~\ref{fig13}(c) and Fig.~\ref{fig13}(d) respectively.
Since all solutions on the Pareto boundary are non-dominated by each other, the metric's improvement on one slice is accompanied by metrics' degradation on the other one or two slices.
Explicitly, the throughput, average delay and SINR cannot be simultaneously improved, indicating their tradeoff.
For example, when the average SINR of class-3 slices is higher than 7 dB, the average delay of class-2 slices will also exceed 6 ms, and the throughput of class-1 slices will be reduced below 36Mbps.
In practical applications, the MNO may activate one of the near-Pareto optimal solutions by referring to slices' priority, where each solution involves a specific resource allocation.
Suppose that the low-delay slices own a higher priority, MNO may activate a near-Pareto optimal solution having the lowest delay, while guaranteeing basic target requirements of the high-throughput slices and the wide-coverage slices.

\section{Conclusions}
In this paper, we simultaneously support high-throughput slices, low-delay slices and wide-coverage slices under the same 3D physical SAGIN.
In order to jointly optimize the throughput, average delay and SINR, we have formulated a non-scalar MOOP involving the optimal vUAVs' position decisions, dynamic network components' deployment and subchannel as well as power allocation.
The CDMADDPG algorithm has been proposed for solving the non-scalar MOOP and for approaching the Pareto optimal solutions, where the centralized unit was used for the global resource allocation policies among the three classes of slices, while three distributed slice-specific units were constructed for making local resource decisions within the slices.
Moreover, a dual resource allocation mechanism was designed for improving the algorithm's performance and the users' QoS.
The proposed algorithm exhibited better performance than the traditional MADDPG algorithm in multi-user scenarios despite its lower computational complexity.
The proposed scheme has also exhibited prominent performance advantages over the single resource allocation mechanism and the fixed vUAV position based scheme.
Compared to the traditional scalar method, our scheme has the distinct benefit of operating without designing the utility function and weights in advance, while still finding numerous near-Pareto optimal solutions,
characterizing the set of tradeoff amongst the different slices.

\appendices
\section{Non-dominance Proof of \eqref{45b}}
Firstly, we assume that the four-tuple found by maximal $r_1^t$ ($\chi _{}^*$) can be dominated by other four-tuples in the space of $\bf{D_1}$.
By \textit{Definition 1} of Pareto optimal solutions in \cite{52}, there must be another solution of $\left\langle \! {{\bf{o}}_{\bf{1}}^{j},\!{\bf{a}}_{\bf{i}}^{j},\!r_{i}^{j},\!{\bf{o'}}_{\bf{1}}^{j}} \! \right\rangle$ associated with $r_{d-s}^j \geq r^*_{d-s},s\!=\!1,2,3$.
In this way, the index $\bf{{\chi}^j_s}$ by $r_{d-s}^j$ must be higher than the index $\bf{{\chi}^*_s}$ by $r^*_{d-s}$,
and then the sum of the three agents' indices $\bf{{\chi}^j}$ must be higher than the sum of the three agents' indices $\bf{{\chi}^*}$.

However, we know that $\bf{{\chi}^*}$ is the maximum sum of indices.
That is, there exists the contradiction between $\bf{{\chi}^*}$ and $\bf{{\chi}^j}$, which indicates the assumption is false.
Therefore, the four-tuple $\left\langle \! {{\bf{o}}_{\bf{1}}^{*},\!{\bf{a}}_{\bf{i}}^{*},\!r_{i}^{*},\!{\bf{o'}}_{\bf{1}}^{*}} \! \right\rangle$ found by maximal $r_1^t$ is not dominated by any other four-tuples in the space of $\bf{D_1}$.

%
%
%
%
%
%
%
%




\begin{thebibliography}{1}
\bibitem{1a}
D. C. Nguyen \textit{et al.}, ``6G Internet of Things: A Comprehensive Survey," \textit{IEEE Internet of Things Journal}, vol. 9, no. 1, pp. 359-383, 1 Jan., 2022.
\bibitem{1}
Q. Wu \textit{et al.}, ``A Comprehensive Overview on 5G-and-Beyond Networks With UAVs: From Communications to Sensing and Intelligence," \textit{IEEE Journal on Selected Areas in Communications}, vol. 39, no. 10, pp. 2912-2945, Oct. 2021.
\bibitem{2}
X. You, CX. Wang, J. Huang \textit{et al.}, ``Towards 6G wireless communication networks: vision, enabling technologies, and new paradigm shifts," \textit{Sci. China Inf. Sci.}, vol. 64, pp. 110301, Jan. 2021.
\bibitem{3}
Z. Jia, M. Sheng, J. Li and Z. Han, ``Toward Data Collection and Transmission in 6G Space-Air-Ground Integrated Networks: Cooperative HAP and LEO Satellite Schemes," \textit{IEEE Internet of Things Journal}, vol. 9, no. 13, pp. 10516-10528, 1 July, 2022.
\bibitem{4}
J. Liu, X. Du, J. Cui, M. Pan and D. Wei, ``Task-Oriented Intelligent Networking Architecture for the Space-Air-Ground-Aqua Integrated Network," \textit{IEEE Internet of Things Journal}, vol. 7, no. 6, pp. 5345-5358, June 2020.
\bibitem{50}
J. Zhang, \textit{et al.}, ``Aeronautical ad-hoc networking for the Internet-above-the-clouds," \textit{Proceedings of the IEEE}, vol. 107, no. 5, pp. 868-911, May 2019.
\bibitem{777}
Y. Wang, Z. Su, J. Ni, N. Zhang and X. Shen, ``Blockchain-Empowered Space-Air-Ground Integrated Networks: Opportunities, Challenges, and Solutions," \textit{IEEE Communications Surveys \& Tutorials}, vol. 24, no. 1, pp. 160-209, Firstquarter 2022.
\bibitem{23}
F. Lyu \textit{et al.}, ``Service-Oriented Dynamic Resource Slicing and Optimization for Space-Air-Ground Integrated Vehicular Networks," \textit{IEEE Transactions on Intelligent Transportation Systems}, vol. 23, no. 7, pp. 7469-7483, July 2022.
\bibitem{10}
H. Cao \textit{et al.}, ``Toward Tailored Resource Allocation of Slices in 6G Networks With Softwarization and Virtualization," \textit{IEEE Internet of Things Journal}, vol. 9, no. 9, pp. 6623-6637, 1 May, 2022.
\bibitem{11}
I. Afolabi, T. Taleb, K. Samdanis, A. Ksentini and H. Flinck, ``Network Slicing and Softwarization: A Survey on Principles, Enabling Technologies, and Solutions," \textit{IEEE Communications Surveys \& Tutorials}, vol. 20, no. 3, pp. 2429-2453, thirdquarter 2018.
\bibitem{11-1}
D. Wu, Z. Zhang, S. Wu, J. Yang and R. Wang, ``Biologically Inspired Resource Allocation for Network Slices in 5G-Enabled Internet of Things," \textit{IEEE Internet of Things Journal}, vol. 6, no. 6, pp. 9266-9279, Dec. 2019.
\bibitem{11-2}
P. Yang, X. Xi, T. Q. S. Quek, J. Chen, X. Cao and D. Wu, ``RAN Slicing for Massive IoT and Bursty URLLC Service Multiplexing: Analysis and Optimization," \textit{IEEE Internet of Things Journal}, vol. 8, no. 18, pp. 14258-14275, 15 Sept., 2021.
\bibitem{12aa}
Z. Fei, B. Li, S. Yang, C. Xing, H. Chen and L. Hanzo, ``A Survey of Multi-Objective Optimization in Wireless Sensor Networks: Metrics, Algorithms, and Open Problems," \textit{IEEE Communications Surveys \& Tutorials}, vol. 19, no. 1, pp. 550-586, Firstquarter 2017.
\bibitem{37}
E. Bjornson, E. Jorswieck, M. Debbah, \textit{et al.}, ``Multi-Objective Signal Processing Optimization: The Way to Balance Conflicting Metrics in 5G systems," \textit{IEEE Signal Processing Magazine}, vol. 31, no. 6, pp. 14-23, 2014.
\bibitem{12}
Q. Shi, L. Zhao, Y. Zhang, G. Zheng, F. R. Yu and H. Chen, ``Energy-Efficiency Versus Delay Tradeoff in Wireless Networks Virtualization," \textit{IEEE Transactions on Vehicular Technology}, vol. 67, no. 1, pp. 837-841, Jan. 2018.
\bibitem{13}
I. Afolabi, J. Prados-Garzon, M. Bagaa, T. Taleb and P. Ameigeiras, ``Dynamic Resource Provisioning of a Scalable E2E Network Slicing Orchestration System," \textit{IEEE Transactions on Mobile Computing}, vol. 19, no. 11, pp. 2594-2608, 1 Nov. 2020.
\bibitem{14}
G. Zhou, L. Zhao, K. Liang, G. Zheng and L. Hanzo, ``Utility Analysis of Radio Access Network Slicing," \textit{IEEE Transactions on Vehicular Technology}, vol. 69, no. 1, pp. 1163-1167, Jan. 2020.
\bibitem{15}
G. Wang, G. Feng, T. Q. S. Quek, S. Qin, R. Wen and W. Tan, ``Reconfiguration in Network Slicing-Optimizing the Profit and Performance," \textit{IEEE Transactions on Network and Service Management}, vol. 16, no. 2, pp. 591-605, June 2019.
\bibitem{16}
R. Lowe, Y. I. Wu, A. Tamar, J. Harb, O. P. Abbeel, and I. Mordatch, ``Multi-agent actor-critic for mixed cooperative-competitive environments," \textit{Advances in neural information processing systems}, pp. 6379-6390, 2017.
\bibitem{17}
J. Cui, S. X. Ng, D. Liu, J. Zhang, A. Nallanathan and L. Hanzo, ``Multiobjective Optimization for Integrated Ground-Air-Space Networks: Current Research and Future Challenges," \textit{IEEE Vehicular Technology Magazine}, vol. 16, no. 3, pp. 88-98, Sept. 2021.
\bibitem{18}
J. Wang, C. Jiang, H. Zhang, Y. Ren, K. -C. Chen and L. Hanzo, ``Thirty Years of Machine Learning: The Road to Pareto-Optimal Wireless Networks," \textit{IEEE Communications Surveys \& Tutorials}, vol. 22, no. 3, pp. 1472-1514, thirdquarter 2020.
\bibitem{20a}
B. Gu, X. Yang, Z. Lin, W. Hu, M. Alazab and R. Kharel, ``Multiagent Actor-Critic Network-Based Incentive Mechanism for Mobile Crowdsensing in Industrial Systems," \textit{IEEE Transactions on Industrial Informatics}, vol. 17, no. 9, pp. 6182-6191, Sept. 2021.
\bibitem{20b}
Y. Zhang, Z. Zhuang, F. Gao, J. Wang and Z. Han, ``Multi-Agent Deep Reinforcement Learning for Secure UAV Communications," 2020 IEEE Wireless Communications and Networking Conference (WCNC), 2020, pp. 1-5.
\bibitem{12.2}
G. Gui, M. Liu, F. Tang, N. Kato, and F. Adachi, ``6G: Opening New Horizons for Integration of Comfort, Security, and Intelligence," \textit{IEEE Wireless Communications}, vol. 27, no. 5, pp. 126-132, Oct. 2020.
\bibitem{5}
J. Liu, Y. Shi, Z. M. Fadlullah and N. Kato, ``Space-Air-Ground Integrated Network: A Survey," \textit{IEEE Communications Surveys \& Tutorials}, vol. 20, no. 4, pp. 2714-2741, Fourthquarter 2018.
\bibitem{12.1}
J. Sun, F. Liu, Y. Zhou, G. Gui, T. Ohtsuki, S. Guo, and F. Adachi, ``Surveillance Plane Aided Air-Ground Integrated Vehicular Networks: Architectures, Applications, and Potential," \textit{IEEE Wireless Communications}, vol. 27, no. 6, pp. 122-128, Dec. 2020.
\bibitem{6}
Y. Wang \textit{et al.}, ``Joint Resource Allocation and UAV Trajectory Optimization for Space-Air-Ground Internet of Remote Things Networks," \textit{IEEE Systems Journal}, vol. 15, no. 4, pp. 4745-4755, Dec. 2021.
\bibitem{7}
X. Cao, B. Yang, C. Yuen and Z. Han, ``HAP-Reserved Communications in Space-Air-Ground Integrated Networks," \textit{IEEE Transactions on Vehicular Technology}, vol. 70, no. 8, pp. 8286-8291, Aug. 2021.
\bibitem{8}
S. Mao, S. He and J. Wu, ``Joint UAV Position Optimization and Resource Scheduling in Space-Air-Ground Integrated Networks With Mixed Cloud-Edge Computing," \textit{IEEE Systems Journal}, vol. 15, no. 3, pp. 3992-4002, Sept. 2021.
\bibitem{9}
J. Ye, S. Dang, B. Shihada and M. -S. Alouini, ``Space-Air-Ground Integrated Networks: Outage Performance Analysis," \textit{IEEE Transactions on Wireless Communications}, vol. 19, no. 12, pp. 7897-7912, Dec. 2020.
\bibitem{16a}
D. A. V. Veldhuizen and G. B. Lamont, ``Multiobjective Evolutionary Algorithms: Analyzing the State-of-the-Art," \textit{Evolutionary Computation}, vol. 8, no. 2, pp. 125-147, June 2000.
\bibitem{19}
H. D. Chantre and N. L. Saldanha da Fonseca, ``The Location Problem for the Provisioning of Protected Slices in NFV-Based MEC Infrastructure," \textit{IEEE Journal on Selected Areas in Communications}, vol. 38, no. 7, pp. 1505-1514, July 2020.
\bibitem{36a}
J. Cui, H. Yetgin, D. Liu, J. Zhang, S. X. Ng and L. Hanzo, ``Twin-Component Near-Pareto Routing Optimization for AANETs in the North-Atlantic Region Relying on Real Flight Statistics," \textit{IEEE Open Journal of Vehicular Technology}, vol. 2, pp. 346-364, 2021.
\bibitem{21}
X. Foukas, G. Patounas, A. Elmokashfi and M. K. Marina, ``Network Slicing in 5G: Survey and Challenges," \textit{IEEE Communications Magazine}, vol. 55, no. 5, pp. 94-100, May 2017.
\bibitem{26}
S. Zhang, H. Zhang, B. Di and L. Song, ``Cellular UAV-to-X Communications: Design and Optimization for Multi-UAV Networks," \textit{IEEE Transactions on Wireless Communications}, vol. 18, no. 2, pp. 1346-1359, Feb. 2019.
\bibitem{27}
R. d. Silva and S. Rajasinghege, ``Optimal Desired Trajectories of UAVs in Private UAV Networks," 2018 International Conference on Advanced Technologies for Communications (ATC), 2018, pp. 310-314.
\bibitem{32}
Y. Shi, Y. Xia and Y. Gao, ``Joint Gateway Selection and Resource Allocation for Cross-Tier Communication in Space-Air-Ground Integrated IoT Networks," \textit{IEEE Access}, vol. 9, pp. 4303-4314, 2021.
\bibitem{31}
M. Samir, S. Sharafeddine, C. M. Assi, T. M. Nguyen and A. Ghrayeb, ``UAV Trajectory Planning for Data Collection from Time-Constrained IoT Devices," \textit{IEEE Transactions on Wireless Communications}, vol. 19, no. 1, pp. 34-46, Jan. 2020.
\bibitem{24}
S. Asmussen, \textit{Applied Probability and Queues}. New York, USA: Springer-Verlag, 2003.
\bibitem{25}
G. Wang, S. Zhou and Z. Niu, ``Radio Resource Allocation for Bidirectional Offloading in Space-Air-Ground Integrated Vehicular Network," \textit{Journal of Communications and Information Networks}, vol. 4, no. 4, pp. 24-31, Dec. 2019.
\bibitem{32c}
Y. Cao, S. -Y. Lien, Y. -C. Liang, K. -C. Chen and X. Shen, ``User Access Control in Open Radio Access Networks: A Federated Deep Reinforcement Learning Approach," \textit{IEEE Transactions on Wireless Communications}, vol. 21, no. 6, pp. 3721-3736, June 2022.
\bibitem{3.3}
D. Liu, J. Zhang, J. Cui, S.-X. Ng, R. G. Maunder, and L. Hanzo, ``Deep Learning Aided Routing for Space-Air-Ground Integrated Networks Relying on Real Satellite, Flight, and Shipping Data," \textit{IEEE Wireless Communications}, vol. 29, no. 2, pp. 177-184, Apr. 2022.
\bibitem{34}
T. Brys, \textit{et al.}, ``Multi-objectivization and Ensembles of Shapings in Reinforcement Learning," \textit{Neurocomputing}, vol. 263, pp. 48-59, Nov. 2017.
\bibitem{35}
D. Silver, J. Schrittwieser, K. Simonyan, \textit{et al.}, ``Mastering the game of Go without human knowledge," \textit{Nature}, vol. 550, pp. 354-359, Ooc. 2017.
\bibitem{36}
G. Sun, Z. T. Gebrekidan, G. O. Boateng, D. Ayepah-Mensah and W. Jiang, ``Dynamic Reservation and Deep Reinforcement Learning Based Autonomous Resource Slicing for Virtualized Radio Access Networks," \textit{IEEE Access}, vol. 7, pp. 45758-45772, 2019.
\bibitem{36c}
J. Zhang, D. Liu, S. Chen, S. X. Ng, R. G. Maunder and L. Hanzo, ``Multiple-Objective Packet Routing Optimization for Aeronautical Ad-Hoc Networks," \textit{IEEE Transactions on Vehicular Technology}, vol. 72, no. 1, pp. 1002-1016, Jan. 2023.
\bibitem{36b}
J. Mei, X. Wang, K. Zheng, G. Boudreau, A. B. Sediq and H. Abou-Zeid, ``Intelligent Radio Access Network Slicing for Service Provisioning in 6G: A Hierarchical Deep Reinforcement Learning Approach," \textit{IEEE Transactions on Communications}, vol. 69, no. 9, pp. 6063-6078, Sept. 2021.
\bibitem{42}
Y. Gao, S. Yang, F. Li, S. Trajanovski, P. Zhou, P. Hui and X. Fu, ``Video Content Placement at the Network Edge: Centralized and Distributed Algorithms," \textit{IEEE Transactions on Mobile Computing}, DOI 10.1109/TMC.2022.3200401.
\bibitem{43}
J. Su, S. Adams and P. A. Beling, ``Value-Decomposition Multi-Agent Actor-Critics," Thirty-Fifth AAAI Conference on Artificial Intelligence (AAAI21), May 2021.
\bibitem{44}
V. R. Konda and J. N. Tsitsiklis, ``Actor-critic algorithms," \textit{Advances in neural information processing systems}, 1008-1014, 2000.
\bibitem{49}
C. Zhou, \textit{et al.}, ``Deep Reinforcement Learning for Delay-Oriented IoT Task Scheduling in SAGIN," \textit{IEEE Transactions on Wireless Communications}, vol. 20, no. 2, pp. 911-925, Feb. 2021.
\bibitem{52}
G. Zhou, L. Zhao, G. Zheng, Z. Xie, S. Song and K. -C. Chen, ``Joint Multi-Objective Optimization for Radio Access Network Slicing Using Multi-Agent Deep Reinforcement Learning," \textit{IEEE Transactions on Vehicular Technology}, vol. 72, no. 9, pp. 11828-11843, Sept. 2023.
\end{thebibliography}
\end{document}